# Can artificial neural networks supplant the polygene risk score for risk prediction of complex disorders given very large sample sizes?

Carlos Pinto (1), Michael Gill (1), Schizophrenia Working Group of the Psychiatric Genomics Consortium (2), Elizabeth A. Heron (1)

1. *Department of Psychiatry, Trinity College Dublin, Ireland.*
2. *Members and their affiliations appear in the Supporting Information.*

**Corresponding author*

*E-mail: eaheron@tcd.ie*

Genome-wide association studies (GWAS) provide a means of examining the common genetic variation underlying a range of traits and disorders. In addition, it is hoped that GWAS may provide a means of differentiating affected from unaffected individuals. This has potential applications in the area of risk prediction. Current attempts to address this problem focus on using the polygene risk score (PRS) to predict case-control status on the basis of GWAS data. However this approach has so far had limited success for complex traits such as schizophrenia (SZ). This is essentially a classification problem. Artificial neural networks (ANNs) have been shown in recent years to be highly effective in such applications. Here we apply an ANN to the problem of distinguishing SZ patients from unaffected controls. We compare the effectiveness of the ANN with the PRS in classifying individuals by case-control status based only on genetic data from a GWAS. We use the schizophrenia dataset from the Psychiatric Genomics

Consortium (PGC) for this study. Our analysis indicates that the ANN is more sensitive to sample size than the PRS. As larger and larger sample sizes become available, we suggest that ANNs are a promising alternative to the PRS for classification and risk prediction for complex genetic disorders.

## Background

Genome-wide association studies (GWAS) capture common genetic variation in the form of single nucleotide polymorphisms (SNPs)and hence provide a means of investigating the genetic architecture of common heritable traits and disorders. In particular, the case-control GWAS, the most frequent GWAS design, has been used to identify systematic genetic differences between cases and unaffected controls in many complex disorders such as schizophrenia (SZ) [17]. One can also consider the inverse problem; that is, given the genetic profile of an individual, can we predict their case or control status? This has obvious applications in risk prediction in the clinical and public health areas even in the absence of a detailed understanding of the genetic mechanisms involved. But the classification problem is still a challenge. This is due to small effect sizes and the random variation of SNP allele frequencies as a result of finite sample sizes, as well as the polygenic nature of complex genetic disorders.

This classification problem has been addressed (in the context of GWAS) principally through the use of the polygenic risk score (PRS) ([20], [8], [10]). For a PRS analysis, each individual in a test dataset is assigned a PRS which summarises their risk for the disorder or trait in question. The PRS is calculated by summing up the number of risk alleles the individual possesses for each SNP weighted by its effect size obtained from an independent *training* GWAS association analysis. The SNPs are all treated as independent. Examples are beginning to appear of the use of the PRS in a clinical setting. For example, the PRS has shown potential for screening individuals for breast cancer risk [13] but its clinical utility is currently limited [26]. For complex disorders such as SZ, the proportion of liability explained by the PRS is relatively low (about 7% for SZ, [17]). Thus, the PRS on its own does not have the ability to discriminate effectively between affected and unaffected individuals [11]. In addition, standard independent testing of SNPs for association with traits is restrictive and does not allow for the possibility of identifying joint associations and interactions between SNPs.

In addition, it has been suggested that, due to the large number of SNPs used in the analysis, that (even after correction for population stratification) a significant fraction of the signal detected by the PRS is actually due to residual population effects [7].

Genetic classification can be viewed as a pattern recognition problem; a class of problems for which artificial neural networks (ANNs) are particularly well suited. ANNs constitute a specific form of machine learning. The possibility of using machine learning methods for classification in genetics has been considered for over a decade (see, for example [18], [21], [22] for some early reviews). More recently, attention has been drawn to the potential of deep learning, essentially a form of ANN using large numbers of hidden layers [27]. GWAS data present particular challenges for methods such as ANNs, in that the number of features (SNPS) is typically large in comparison with the number of training examples (sample size). Early studies (see, for example [1], [2], [3], [4], [5]), used small sample sizes (typically a few hundred samples) and in fact only a few tens of SNPs, although clinical covariates were typically included as well. The number of SNPs (9000) used in a recent study ([6]) was comparable to the number used in the work we report here but the sample size was still small (approximately 1800). Here we exploit the large sample sizes now available to investigate the potential of ANNs to predict disease risk on the basis of genetic data alone given a sufficient number of training examples. The use of genetic data alone potentially allows the prediction of risk before disease onset when clinical data are not yet available. The specific dataset that we use is the SZ dataset from the Psychiatric Genomics Consortium (PGC), (containing approximately 64,000 samples), see Methods section.

## Results

### Sensitivity to Number of SNPs

The results of our analysis are shown in Fig 1 and Table 1 where we have varied the number of SNPs, for different sample sizes (10,000, 30,000, and 64,542). For each sample size we show the show the variation in performance for different p-value thresholds. Classification efficiency is measured as area under the curve (AUC). The ANN performs better than the PRS when small numbers of SNPs are used in the analysis. However, both methods become comparable as the number of SNPs increases for a given sample size (see Fig 1, plots (a), (b), (c)). Both methods show an improvement in performance that levels off as the number of SNPs increases beyond a certain point.

### Sensitivity to Sample Size

Results for our sensitivity to sample size analysis can be seen in Fig 2 and Table 2 where we have varied the sample size for three different numbers of SNPs. Our results indicate a higher level of performance for the ANN relative to the PRS in Fig 2, plots (a) and (b). This difference in performance increases with increasing sample size. In Fig 2, plot (c) the ANN performs less well for lower sample sizes (10,000, 20,000, and 30,000). For larger sample sizes it outperforms the PRS.

## Discussion

We have carried out analyses on the performance of an ANN and the PRS in classifying case-control status using GWAS schizophrenia data. We conducted sensitivity analyses with respect to both number of SNPs and sample size. Before discussing the results, we briefly review the differences between these two approaches to classification. In the PRS approach, an individual in a test dataset is assigned case or control status according to their risk score. The score is calculated by summing the number of risk alleles the individual possesses for each SNP weighted by its effect size. This effect size is obtained from an independent (training) dataset. The SNPs are treated as independent and the score is refined by increasing the number of SNPs in the analysis. The precision with which the effect size is estimated is improved by increasing the sample size in the training dataset. In the ANN, on the other hand, all SNPs are simultaneously taken into account for each sample. The network weights are adjusted accordingly, using the data from the training set and these weights are used to calculate probabilities of class membership in a test dataset. The probabilities are refined by increasing the sample size in the training dataset.

These two approaches are therefore complementary, and while we would expect the performance of both to increase with both sample size and number of SNPs, the behaviour of the two methods as these parameters vary will, in general, differ.

In our sensitivity to number of SNPs analysis (Fig 1, Table 1) the ANN initially performs better than the PRS. However, the PRS improves more quickly as the number of SNPs increases for a given sample size (see Fig 1, plots (a), (b), (c)). In both cases, the rate of improvement slows as the number of SNPs increases beyond a certain point. We have restricted the analysis to p-value thresholds less than 0.05 (corresponding to approximately 16,000 SNPs for the largest sample size). This is partly for practical reasons (computation time for the ANN). It is also desirable to keep the number of features reasonably small compared to the number of training examples. It is possible that a further improvement in performance may result if the number of SNPs is greatly increased. Our objective here is not to achieve the maximum possible performance but to compare the behaviour of the two methods as the number of SNPs is varied. Performance, on the other hand, does improve as the sample size is increased (for example, at a p-value threshold of 0.05 with overall sample size of 10,000, mean (se) ANN AUC = 0.578 (0.006), mean (se) PRS AUC = 0.584 (0.005), mean (sd) number of SNPs = 9,681 (124), with overall sample size 30,000, mean (se) ANN AUC = 0.617 (0.004), mean (se) PRS AUC = 0.628 (0.004), mean (sd) number of SNPs = 12,184 (109), and with overall sample size of 64,542, mean (se) ANN AUC = 0.661 (0.002), mean (se) PRS AUC = 0.669 (0.002), mean (sd) number of SNPs = 16,054 (154), see Fig 1, plots (a), (b), (c) and Table 1).

Note that the same p-value threshold at the larger sample sizes (for example, Fig 1, plot (c) in comparison to Fig 1 plot (a)) implies an increase in the number of SNPs available for analysis. This means that we cannot make a definitive statement about the behavior of the methods as sample size increases on the basis of the SNP sensitivity analysis above. This issue is addressed explicitly in our second sensitivity analysis where the number of SNPs is kept approximately constant as we vary the sample size.

Results for our sensitivity to sample size analysis can be seen in Fig 2 and Table 2 where we have varied the sample size for three different sets of SNPs obtained from different p-value thresholds in a reference dataset. The maximum threshold used was 0.015 corresponding to approximately 3500 SNPs. The reason for this restriction was computation time for the ANN. It is likely that performance will improve further if a larger number of SNPs is used, however our objective here was to compare the behaviour of the two methods as sample size is varied, rather than to achieve optimum performance. Our results indicate a higher level of performance for the ANN relative to the PRS in Fig 2, plot (a) with 661 SNPs (mean) and plot (b) with 1,377 SNPs (mean). This difference in performance increases with increasing sample size. In Fig 2, plot (c), with 3,531 SNPs (mean), the ANN performs more poorly for lower sample sizes (10,000, 20,000, and 30,000). For larger sample sizes it begins to outperform the PRS.

## Conclusions

there are two basic parameters that we can consider when we are carrying out this type of analysis: the number of SNPs and the sample size. We find that the performance of the PRS is more sensitive to the number of SNPS, while the ANN is more sensitive to the sample size. We emphasize that the ANN architecture used here has been minimal (one hidden layer, two hidden nodes). In particular, the architecture has not been specifically optimised for this dataset and our results therefore have general validity. There is considerable scope for further development with regard to the optimisation of the architecture in order to further improve the ANN's performance on this dataset.

Performance in terms of computation time has also not been optimised (see Supporting Information (SI) for details) since we do not regard this as a critical issue at this stage of development. Training times are therefore relatively long, in contrast to the PRS which requires only an association analysis on the training dataset to obtain the odds ratios required to compute the scores. Significant reductions in computation time should be achievable by optimisation of for example, step-size and (in particular) parallelisation.

An important issue for ANNs (and machine learning methods in general) is the dissemination of the results. (This is straightforward in the case of the PRS, since all that is required is the list of SNPs used and the associated odds ratios.) All the information for the trained network is available, in principle, in the list of SNPs and the list of network weights and biases; however this information is difficult to use and interpret in practice and it is desirable to have a complete end-to-end pipeline available. This is beyond the scope of the present work, but we plan to develop such a pipeline in the next phase of this project.

In summary, we have presented here an alternative potential approach to the current method of choice, the PRS for case-control classification. Our analysis indicates that the ANN outperforms the PRS at large sample sizes and will therefore prove a promising alternative to the PRS for the very large sample sizes that are now becoming available. The results we have presented here are a proof-of-concept. Our ANN requires further optimisation to maximise performance. The system will also require further optimisation for speed in order to handle larger numbers of SNPs. In addition, a fully developed pipeline will be required in order for the system to be easily useable by the wide community. These issues will be addressed in the next phase of this work.

## Methods

### Data

We use the PGC SZ data (Ripke et al. [17]). Of the total of 43 datasets analysed in [17] we have been granted access to 40, representing 84.1% of the total case-control data originally

analysed (67,184 of 79,845 samples), (86.7% of the cases (29,689 of 34,241), 82.2% of the controls (37,495 of 45,604)). Each of these datasets has been imputed using the 1000 Genomes Project reference panel [19] and quality control (QC) has been applied to these individual datasets prior to our accessing the data. We use the datasets that have been subjected to a light QC (best guess genotypes based on imputation, SNP missing rate <2%) in order to maximise the overlap of SNPs between the datasets as we are performing a mega analysis rather than a meta-analysis. These datasets have approximately 7,000,000 variants each, see SI Table S1. These data have been made available by the PGC, see Acknowledgements for details of how to access these data.

## Data Quality Control, Filtering and Principal Component Analysis

We merged all 40 datasets using the PLINK 1.9 software [14] command -merge-list. This resulted in a dataset consisting of approximately 14,000,000 variants and 67,184 samples (29,689 cases, 37,495 controls). Next, non-SNP variants were removed, SNP missingness (>0.01) and individual missingness (>0.02) filtering was carried out, and related samples were removed (random removal of one individual from pairs of related samples with cases preferentially retained). Additional SNP QC was then carried out. This resulted in what we refer to as the *Mega* dataset consisting of approximately 2,000,000 variants (not all imputed SNPs were present in all 40 datasets). The dataset consisted of 64,542 samples (28,707 cases, 35,835 controls), see Table 3. This *Mega* dataset was used as the basis for all subsequent analyses. In addition to filtering steps for SNPs and samples being carried out as part of the QC it was necessary to conduct principal component analyses (PCAs) in order to adjust association analyses for population stratification that is present in this merged dataset. For a PCA the dataset was pruned using Plink 1.9 [14] (indep-pairwise 50 5 0.1) and regions of known long-range high linkage disequilibrium (LD) were removed [16]. The PCA analysis was then carried out using the software SmartPCA [15] on the resulting pruned dataset.

## Association Findings for Mega Datasets

As we are conducting a mega analysis here and Ripke et al. [17] conducted a meta-analysis we carried out a comparison of top significant results between the two approaches for reference. We conducted an association analysis on the *Mega* dataset, including the first 10 PCs. Ripke et al. identified 128 LD independent SNPs that exceeded genome-wide significance (p-value $5x10e^{-8}$). For each of these 128 top hits, an associated locus was defined as the physical region containing SNPs correlated at $r^2 > 0.6$ with each of the 128 index SNPs. Of these 128 top hit regions, 111 are associations on chromosomes 1 to 22 (we are not considering sex chromosomes here) and are bi-allelic SNPs, the other 17 are either on the sex chromosomes (3 on chrX) and/or are not bi-allelic SNPs (14 indels). For each of these 111 regions we identified whether or not we had a SNP included in these regions in our *Mega* dataset, resulting in 95 SNP regions that overlap with Ripke et al. [17] (86%). Of these 95 overlapping regions, we identified 39 (41%) that contained a SNP in our analysis that also had a p-value $\leq 5x10e^{-8}$ in our mega association analysis. See Figs S1-S4 in SI for further details. Broadly speaking our results are consistent with those of Ripke et al. [17].

## Artificial Neural Networks

In this section we briefly review the principles of artificial neural networks, as used for classification. An ANN consists of a number of interconnected neurons, or nodes, each of which processes information and passes it to other nodes in the network. There is an input layer which receives its inputs from the user and an output layer which delivers the outputs. In general, there will be one or more intermediate hidden layers; this is the architecture of the network (see Fig 3). In the course of training, both inputs and outputs are specified and the ANN adjusts the network weights in order to achieve the best fit to the data. In this way the ANN learns to recognise patterns. In our application, the inputs are the SNP genotypes for each individual. There is a single output node - the probability of the case or control status of each individual.

The complexity of the behaviour of the ANN increases with the number of hidden nodes and hidden layers. Since in this study we are primarily interested in the general question of the performance of the ANN on large genetic datasets (and not on optimisation with respect to a particular disorder), we used a simple architecture with a single hidden layer consisting of two hidden nodes. The network is trained on a subset of the cases and controls. After training, the network is tested on an independent subset to assess its accuracy in predicting case/control status. We use the Skynet ANN for all analyses [9]. See SI for further details on the computational performance of this ANN.

## Analysis Plan

All supervised learning classification problems incorporate two distinct steps. The first step, *feature selection*, involves determining the inputs that will be used to compute the classification. The second step, *training*, involves allocating appropriate weights to the selected features by using a training dataset in which both features and class labels are supplied. The trained classifier then uses these weights to classify new, previously unseen instances.

We begin by splitting the data into two non-overlapping subsets: training and test. The majority of the data (approximately 90%) is allocated to the training dataset. We perform additional SNP QC filtering (see Part 2, Table 3) independently on the training and test datasets. We conduct a PCA on the training dataset. We then carry out an association analysis correcting for PCs on the training dataset. We select SNPs (based on the association analysis and LD clumping in order to determine independent signals). These SNPs are the features for the classification step. For classification via the PRS the odds ratios (ORs) from the association analysis on the training dataset are supplied, as well as the SNP genotypes of the test dataset. We use the R statistical package PRSice [23] to compute the PRS. For classification in the ANN, the full genotype information for the selected SNPs for both the training and test datasets is required but not the ORs. This is the main difference in the information that is supplied to these classification methods. Based on the information that has been provided for the PRS a score is calculated for each sample in the test dataset which consists of the genotype data for the selected SNPs. In the case of the ANN the training genotypes are used to compute the network weights using the training data. These weights are then used to calculate class membership probabilities for each of the samples in the test dataset. See Figs S5-S8 in SI for details on the workflow. In order to examine the stability of the methods and to get error bounds we created ten independent training and test datasets as follows. We split the dataset into ten random disjoint subsets. Each of these ten is used as a test dataset with its corresponding training dataset consisting of the other nine datasets combined. This ensures that all test datasets are independent of each other and each is independent of their associated training dataset.

## Comparison of Performance

A standard method for assessing the performance of a classifier is the AUC and we employ this here. All AUCs are calculated in R [24] using the package pROC [25].

## Sensitivity to number of SNPs

The aim here is to compare the performance of both methods as we vary the number of SNPs. We examine the performance for three different sample sizes. We randomly select 10,000 (30,000 or use all of the *Mega* dataset) samples from the *Mega* dataset. We then split this dataset into ten disjoint subsets as described above. Each of these ten subsets is considered in turn as a test dataset with the remaining subsets its corresponding training dataset. This gives us ten replicates and enables us to compute a standard error (se) on the results. The ten test datasets are independent of each other by construction and each test dataset is also independent of its corresponding training dataset. For each replicate we carry out QC on the test and training dataset independently as described above. We then conduct an association analysis on the training dataset to obtain p-values for the SNPs. We then use these p-values to clump the SNPs at different levels of significance with results grouped based on LD (–clump command in plink). This ensures independence of the SNPs. As the significance level increases the number of SNPs yielded also increases allowing us to examine the behavior of the methods as we vary the number of SNPs.

## Sensitivity to Sample Size

The aim here is to compare the performance of both methods as we vary the number of samples. We examine the performance for three different p-value thresholds. In order to maintain approximately the same number of SNPs as we vary the sample size it is necessary to generate a reference set of SNPs from an independent subset of the *Mega* dataset. We randomly select 10,000 samples from the *Mega* dataset. We then conduct QC and PCA and an association analysis on this subset of data. We obtain three reference lists of SNPs at p-value thresholds: 0.002, 0.005, and 0.015 and the resulting SNPs are also clumped to ensure independence of signals. These yield SNP sets of size 677, 1,411, and 3,608, respectively. We then use the remaining 54,542 samples of the *Mega* dataset to create ten disjoint subsets as described previous section. For each replicate we carry out QC on the test and training dataset independently as described above. We then conduct an association analysis on the training dataset. This provides the ORs necessary for the PRS. The SNPs to be used in the analysis are selected from the reference list. Note that not all SNPs in the reference list will be present due to QC, particularly at lower sample sizes, but these variations are small, see Table 2. This allows us to examine the behavior of the methods as we vary the sample size using an approximately fixed number of SNPs.

## Ethics declarations

### Ethics approval and consent to participate

Not applicable.

### Consent for publication

Not applicable.

### Competing interests

Not applicable.

## Availability of data and materials

The data that support the findings of this study are available from the Psychiatric Genomics Consortium [http://www.med.unc.edu/pgc]. Restrictions apply to the availability of these data, permission required from the Psychiatric Genomics Consortium.

## Funding

This work was supported by the Wellcome Trust [200608/Z/16/Z].

## Authors' Contributions

CP and EAH conceived the idea, performed the analysis and wrote the paper. MG provided critical feedback and helped shape the research, analysis and manuscript. Members of the Schizophrenia Working Group of the Psychiatric Genomics Consortium provided access to the data used for the analysis. All listed members of the Schizophrenia Working Group of the Psychiatric Genomics Consortium had the opportunity to comment on the manuscript and approved the manuscript.

## Acknowledgments

Data was made available by the Psychiatric Genomics Consortium (http://www.med.unc.edu/pgc). This work was carried out on the Dutch national e-infrastructure with the support of SURF Cooperative and at the Trinity Centre for High

Performance Computing. This study makes use of data generated by the Wellcome Trust Case-Control Consortium. A full list of the investigators who contributed to the generation of the data is available from www.wtccc.org.uk. Funding for the project was provided by the Wellcome Trust under award 076113, 085475 and 090355.

| | *P-value threshold* | *SNPs Mean (sd)* | *Reps* | *ANN Mean (se)* | *PRS Mean (se)* |
|---|---|---|---|---|---|
| ***Sample Size 10,000*** | 0.001 | 368 (19) | 10 | 0.554 (0.004) | 0.549 (0.007) |
| | 0.005 | 1378 (29) | 10 | 0.572 (0.006) | 0.570 (0.006) |
| | 0.010 | 2481 (35) | 10 | 0.584 (0.007) | 0.576 (0.005) |
| | 0.015 | 3481 (53) | 10 | 0.578 (0.006) | 0.579 (0.004) |
| | 0.020 | 4439 (63) | 10 | 0.583 (0.006) | 0.581 (0.005) |
| | 0.035 | 7123 (69) | 10 | 0.586 (0.005) | 0.583 (0.005) |
| | 0.050 | 9681 (124) | 10 | 0.578 (0.006) | 0.584 (0.005) |
| ***Sample Size 30,000*** | 0.001 | 770 (29) | 10 | 0.599 (0.002) | 0.578 (0.003) |
| | 0.005 | 2253 (42) | 10 | 0.611 (0.003) | 0.599 (0.004) |
| | 0.010 | 3686 (46) | 10 | 0.617 (0.003) | 0.610 (0.004) |
| | 0.015 | 4936 (63) | 10 | 0.616 (0.004) | 0.615 (0.004) |
| | 0.020 | 6100 (70) | 10 | 0.617 (0.002) | 0.618 (0.004) |
| | 0.035 | 9286 (106) | 10 | 0.616 (0.004) | 0.623 (0.003) |
| | 0.050 | 12184 (109) | 10 | 0.617 (0.004) | 0.628 (0.004) |
| ***Sample Size 64,542*** | 0.001 | 1727 (20) | 10 | 0.653 (0.001) | 0.625 (0.002) |
| | 0.005 | 3985 (35) | 10 | 0.661 (0.002) | 0.645 (0.002) |
| | 0.010 | 5914 (87) | 10 | 0.661 (0.002) | 0.652 (0.002) |
| | 0.015 | 7536 (108) | 10 | 0.658 (0.002) | 0.655 (0.002) |
| | 0.020 | 8979 (117) | 10 | 0.661 (0.002) | 0.659 (0.002) |
| | 0.035 | 12764 (146) | 10 | 0.659 (0.002) | 0.665 (0.002) |
| | 0.050 | 16054 (154) | 10 | 0.661 (0.002) | 0.669 (0.002) |

**Table 1: Sensitivity to Number of SNPs.** The results presented in this table are also to be found in Fig 1.

| | Sample Size | SNPs Mean (sd) | Reps | ANN Mean (se) | PRS Mean (se) |
|---|---|---|---|---|---|
| ***P-value threshold 0.002*** | | | | | |
| ***~661 SNPs*** | 10000 | 652 (5) | 10 | 0.541 (0.004) | 0.533 (0.007) |
| | 20000 | 655 (4) | 10 | 0.564 (0.004) | 0.552 (0.004) |
| | 30000 | 663 (4) | 10 | 0.576 (0.004) | 0.559 (0.004) |
| | 40000 | 667 (3) | 10 | 0.575 (0.008) | 0.563 (0.003) |
| | 50000 | 665 (3) | 10 | 0.580 (0.002) | 0.561 (0.003) |
| | 54542 | 665 (3) | 10 | 0.583 (0.003) | 0.563 (0.002) |
| ***P-value threshold 0.005*** | | | | | |
| ***~1,377 SNPs*** | 10000 | 1333 (9) | 10 | 0.552 (0.007) | 0.545 (0.005) |
| | 20000 | 1366 (9) | 10 | 0.563 (0.005) | 0.557 (0.004) |
| | 30000 | 1381 (8) | 10 | 0.578 (0.003) | 0.565 (0.003) |
| | 40000 | 1388 (5) | 10 | 0.586 (0.003) | 0.571 (0.003) |
| | 50000 | 1386 (6) | 10 | 0.588 (0.002) | 0.569 (0.003) |
| | 54542 | 1386 (7) | 10 | 0.589 (0.003) | 0.571 (0.002) |
| ***P-value threshold 0.015*** | | | | | |
| ***~3,531 SNPs*** | 10000 | 3469 (24) | 10 | 0.546 (0.011) | 0.553 (0.005) |
| | 20000 | 3508 (15) | 10 | 0.558 (0.003) | 0.568 (0.004) |
| | 30000 | 3539 (18) | 10 | 0.572 (0.003) | 0.578 (0.003) |
| | 40000 | 3557 (11) | 10 | 0.586 (0.002) | 0.581 (0.002) |
| | 50000 | 3554 (13) | 10 | 0.589 (0.004) | 0.585 (0.003) |
| | 54542 | 3557 (11) | 10 | 0.595 (0.003) | 0.584 (0.002) |

**Table 2: Sensitivity to Sample Size.** The results presented in this table are also to be found in Fig 2.

| | | SNPs | Cases | Controls | Samples |
|---|---|---|---|---|---|
| | **Merged Datasets** | | | | |
| | | **13,802,094** | **29,689** | **37,495** | **67,184** |
| **PART 1** | **SNP QC** | | | | |
| | SNPs only, no indels, --snps-only | -865,628 | - | - | - |
| | SNP call rate < 0.99, --geno 0.01 | -10,917525 | - | - | - |
| | **Individual QC** | | | | |
| | Individual call rate < 0.98, --mind 0.02 | - | -216 | -193 | -409 |
| | *Relatedness PI-HAT > 0.2, --genome | - | -766 | -1,467 | -2,233 |
| | **Remaining** | **2,018,941** | **28,707** | **35,835** | **64,542** |
| **PART 2** | **SNP QC** | | | | |
| | SNP call rate < 0.99, --genp 0.01 | -3,650 | - | - | - |
| | Diff missing cases and controls > 0.02 | 0 | - | - | - |
| | HWE – controls <= $10e^{-6}$ | -5,815 | - | - | - |
| | HWE – cases <= $10e^{-10}$ | -16 | - | - | - |
| | MAF < 0.01 | -2,912 | - | - | - |
| **Mega Dataset** | **Remaining** | **2,006,548** | **28,707** | **35,835** | **64,542** |

**Table 3: Quality Control.** Part 1 and Part 2 QC filters applied to the 40 merged datasets. Part 2 filters are applied to the training and test datasets. *Analysis is conducted on a pruned dataset, same pruning applied as for PCA analysis. Pairs of related samples with a PI-HAT >0.2 are identified and one individual is removed from the pair with cases preferentially retained. Analysis is repeated on remaining samples to again check for relatedness.

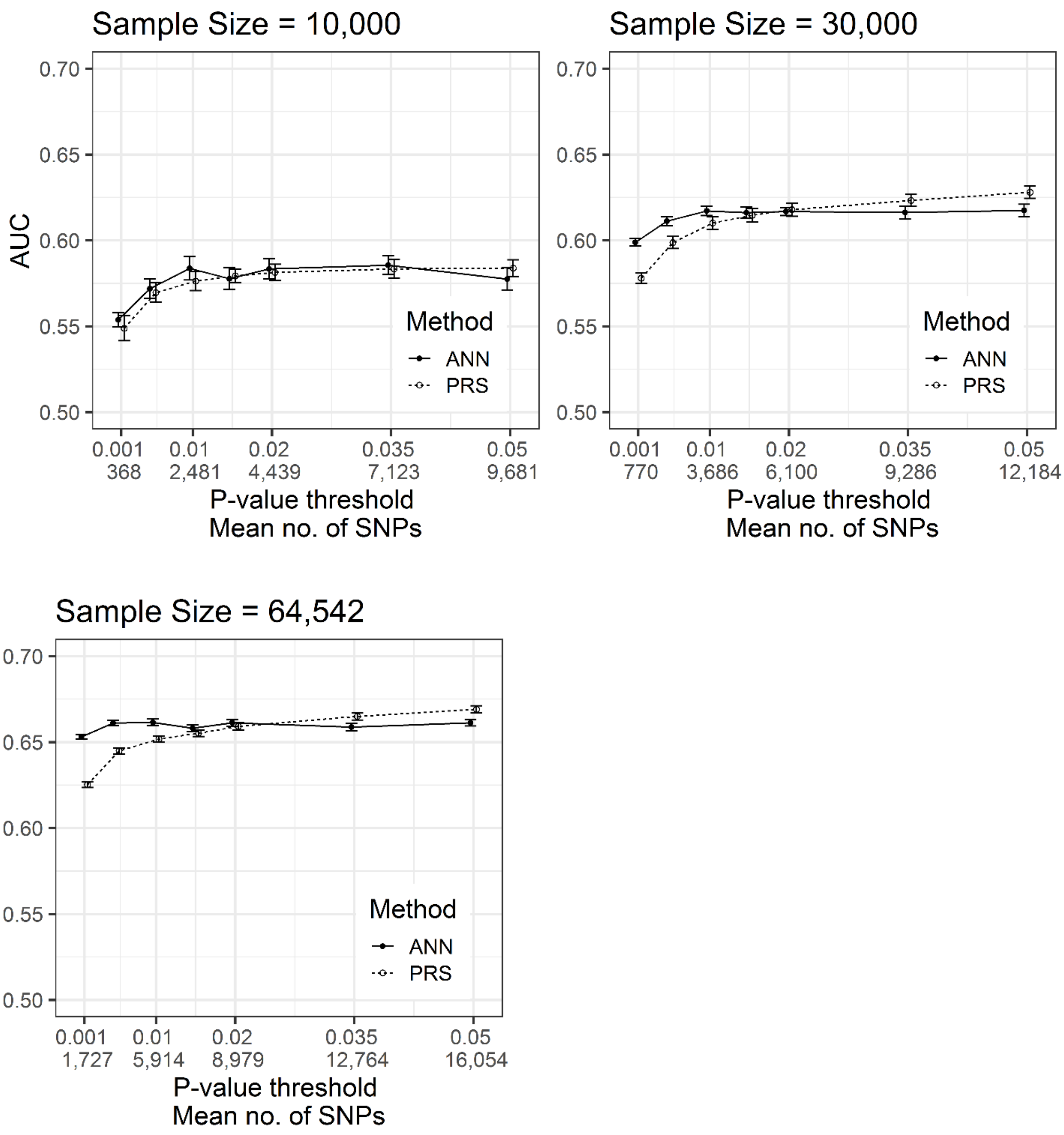

**Figure 1: Sensitivity to Number of SNPs.** The results presented in this figure are also to be found in Table 1.

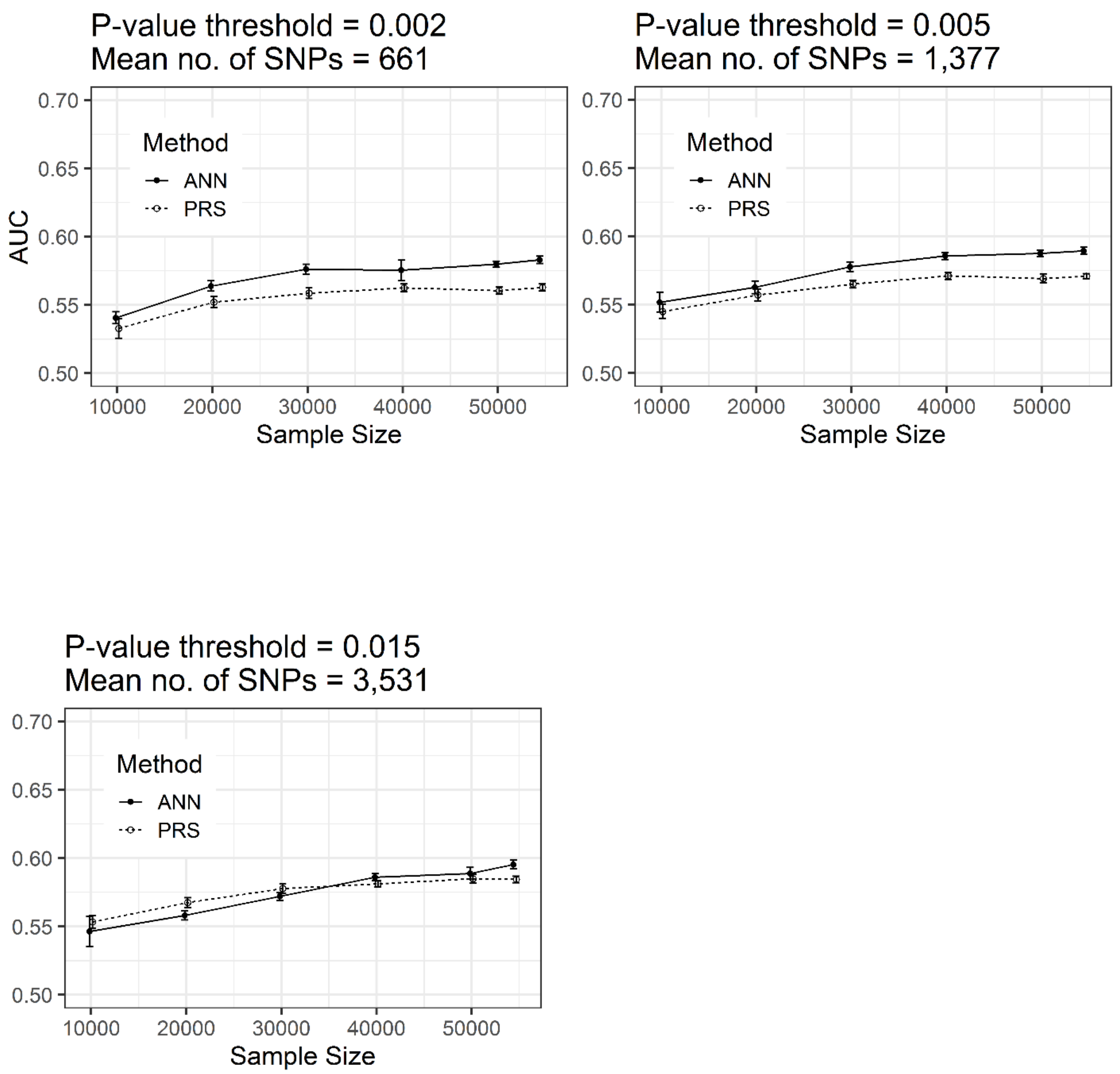

**Figure 2: Sensitivity to Sample Size.** The results presented in this figure are also to be found in Table 2.

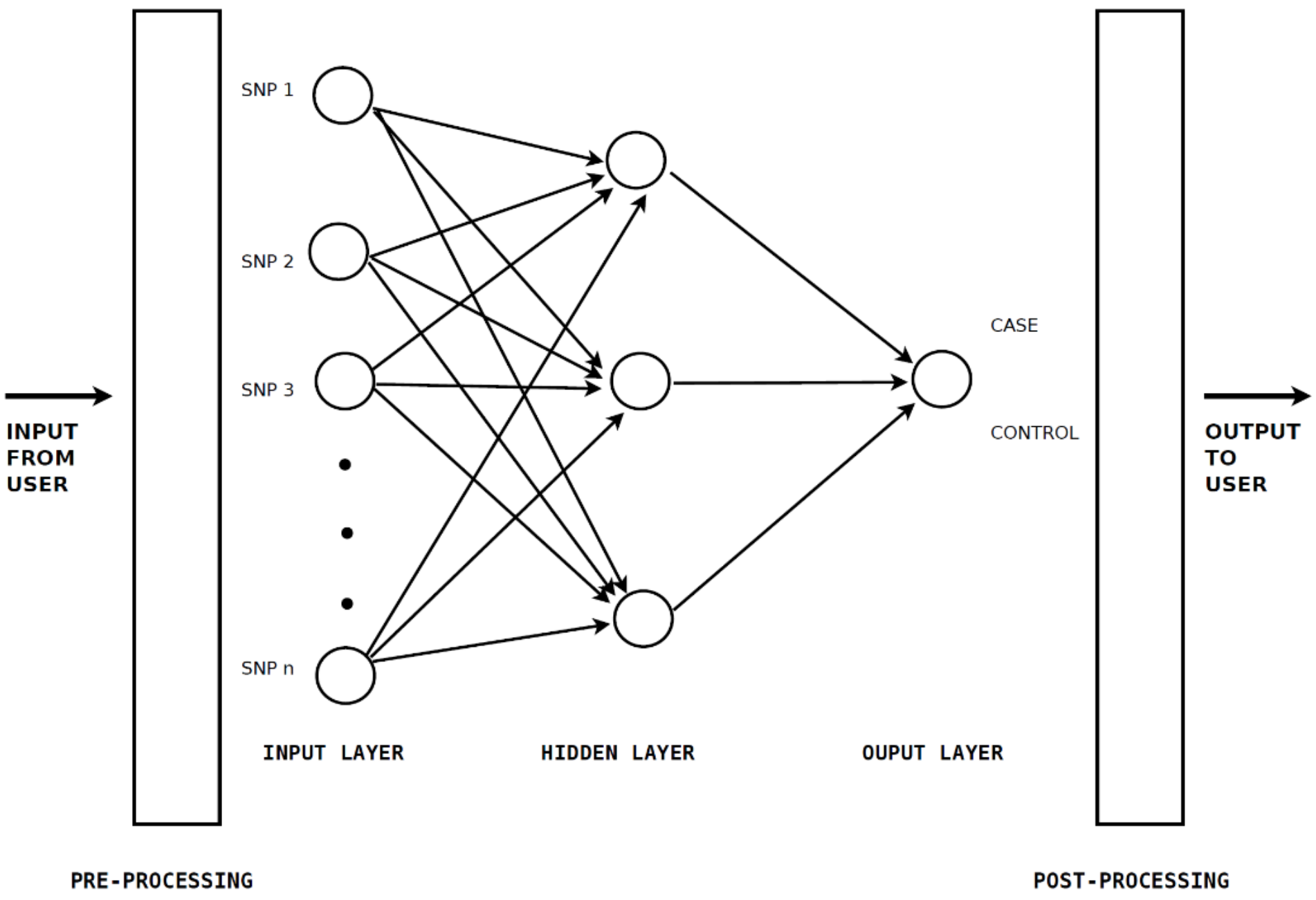

**Figure 3: Artificial Neural Network Architect**

## Supporting Information for:

# Can artificial neural networks supplant the polygene risk score for risk prediction of complex disorders given very large sample sizes?

Carlos Pinto[1], Michael Gill[1], Schizophrenia Working Group of the Psychiatric Genomics Consortium[2], Elizabeth A. Heron[1*]

1. *Department of Psychiatry, Trinity College Dublin, Ireland.*
2. *Members and their affiliations appear in this Supplementary Information.*

**Corresponding author*

*E-mail: eaheron@tcd.ie*

| | Dataset | Cases | Controls | Samples | .bg Variants |
|---|---|---|---|---|---|
| 1 | scz cims eur-qc | 71 | 69 | 140 | 7,176,582 |
| 2 | scz zhh1 eur-qc | 191 | 190 | 381 | 5,204,976 |
| 3 | scz pews eur-qc | 150 | 236 | 386 | 7,292,337 |
| 4 | scz lie2 eur-qc | 137 | 269 | 406 | 7,886,295 |
| 5 | scz swe1 eur-qc | 221 | 214 | 435 | 6,047,632 |
| 6 | scz msaf eur-qc | 327 | 139 | 466 | 7,955,321 |
| 7 | scz port eur-qc | 346 | 216 | 562 | 6,408,457 |
| 8 | scz lacw eur-qc | 157 | 466 | 623 | 7,549,844 |
| 9 | scz edin eur-qc | 368 | 284 | 652 | 7,484,683 |
| 10 | scz ersw eur-qc | 322 | 332 | 654 | 8,132,615 |
| 11 | scz caws eur-qc | 424 | 306 | 730 | 6,122,240 |
| 12 | scz top8 eur-qc | 377 | 403 | 780 | 7,451,253 |
| 13 | scz munc eur-qc | 437 | 351 | 788 | 6,508,392 |
| 14 | scz cati eur-qc | 409 | 392 | 801 | 6,612,133 |
| 15 | scz buls eur-qc | 195 | 608 | 803 | 7,340,965 |
| 16 | scz asrb eur-qc | 509 | 310 | 819 | 7,651,487 |
| 17 | scz lie5 eur-qc | 509 | 389 | 898 | 7,506,231 |
| 18 | scz umes eur-qc | 197 | 713 | 910 | 7,709,209 |
| 19 | scz denm eur-qc | 492 | 458 | 950 | 7,643,941 |
| 20 | scz umeb eur-qc | 375 | 584 | 959 | 8,130,401 |
| 21 | scz uclo eur-qc | 521 | 494 | 1,015 | 5,670,555 |
| 22 | scz 3m eur-qc | 186 | 930 | 1,116 | 7,121,806 |
| 23 | scz dubl eur-qc | 272 | 860 | 1,132 | 7,398,573 |
| 24 | scz cou3 eur-qc | 540 | 693 | 1,233 | 7,781,847 |
| 25 | scz ucla eur-qc | 705 | 637 | 1,342 | 7,642,714 |
| 26 | scz egcu eur-qc | 239 | 1,177 | 1,416 | 7,978,592 |
| 27 | scz aber eur-qc | 720 | 699 | 1,419 | 6,214,060 |
| 28 | scz i6 eur-qc | 361 | 1,082 | 1,443 | 7,933,222 |
| 29 | scz aarh eur-qc | 883 | 873 | 1,756 | 7,809,181 |
| 30 | scz swe6 eur-qc | 1,094 | 1,219 | 2,313 | 8,129,242 |
| 31 | scz gras eur-qc | 1,086 | 1,232 | 2,318 | 7,770,108 |
| 32 | scz irwt eur-qc | 1,309 | 1,022 | 2,331 | 7,521,232 |
| 33 | scz ajsz eur-qc | 896 | 1,595 | 2,491 | 7,794,461 |
| 34 | scz pewb eur-qc | 641 | 1,892 | 2,533 | 7,425,737 |
| 35 | scz boco eur-qc | 1,847 | 2,170 | 4,017 | 7,408,363 |
| 36 | scz clo3 eur-qc | 2,150 | 2083 | 4,233 | 7,961,221 |
| 37 | scz s234 eur-qc | 2,077 | 2,341 | 4,418 | 7,771,960 |
| 38 | scz swe5 eur-qc | 1,801 | 2,617 | 4,418 | 8,061,512 |
| 39 | scz mgs2 eur-qc | 2,681 | 2,653 | 5,334 | 7,543,555 |
| 40 | scz clm2 eur-qc | 3,466 | 4,297 | 7,763 | 7,419,981 |
| | **Total Included*** | **29,595** | **36,205** | **67,184** | |

**Table S1: PGC Datasets Included in Analysis.** *Datasets not available for inclusion in this analysis that were analysed originally in Ripke et al. 2014 [1]: scz jr (includes Johnson and Johnson and Roche cases), scz lktu eu, and scz pa eur. .bg stands for best guess.

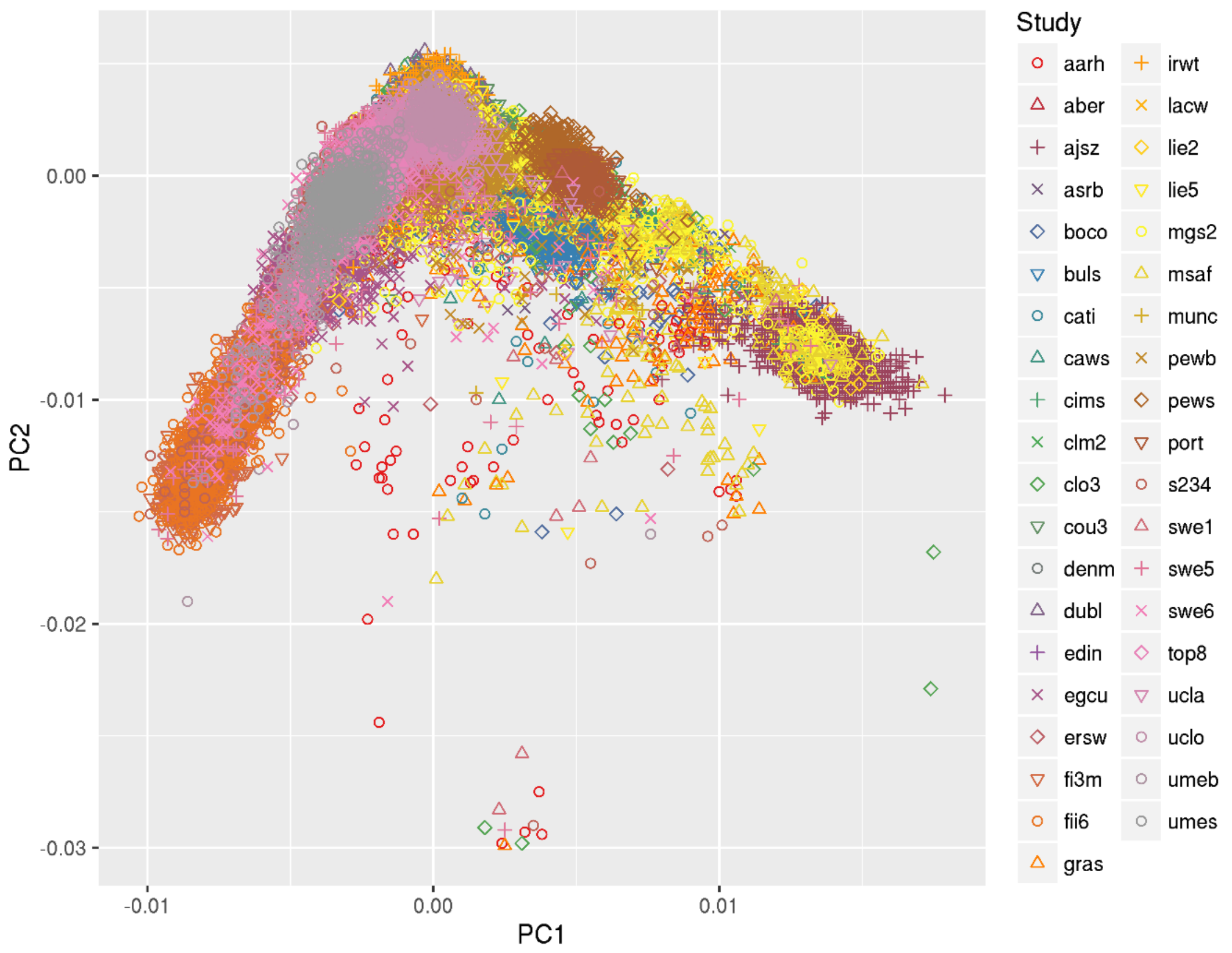

**Figure S1: PC1 vs PC2 for Mega Dataset**

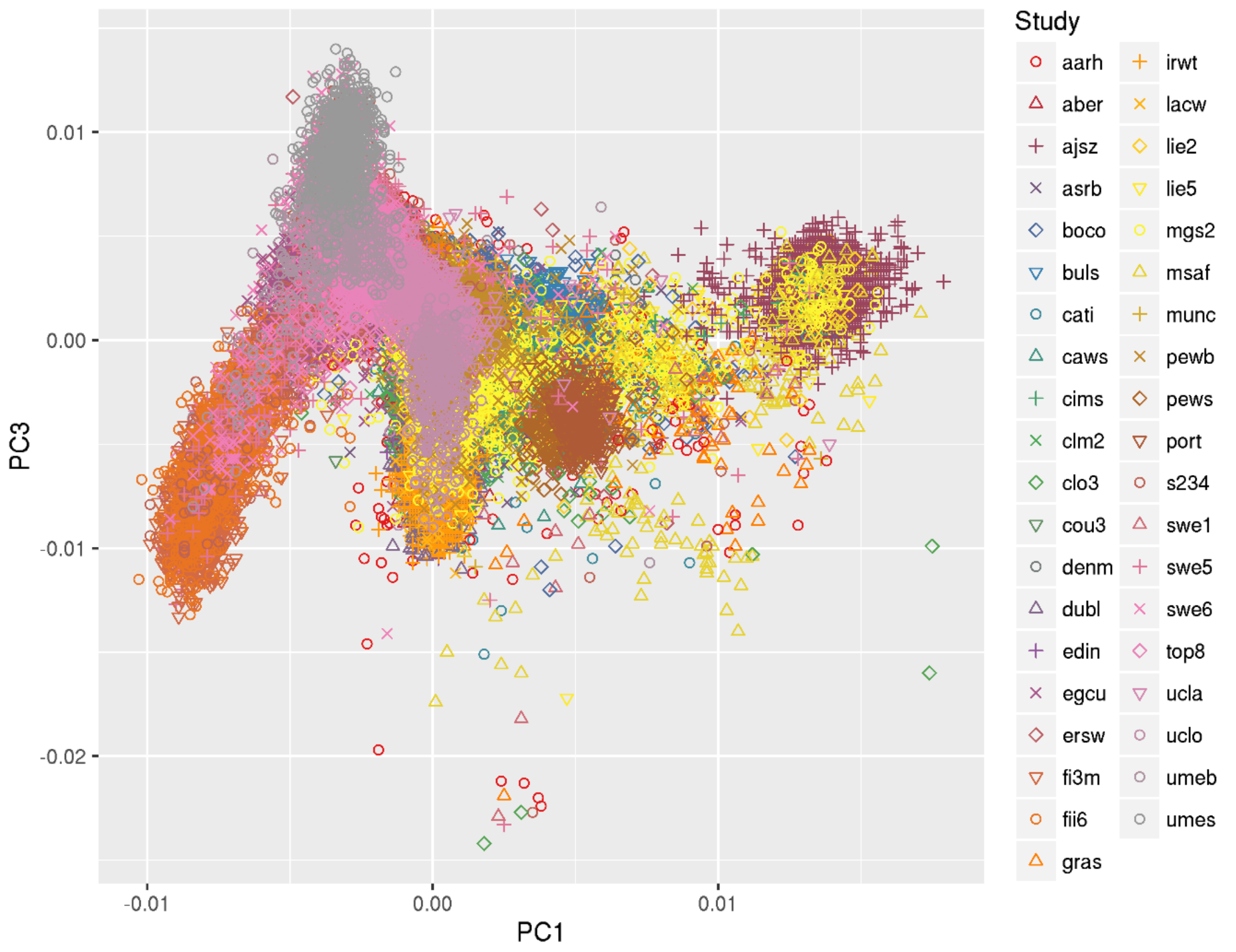

**Figure S2: PC1 vs PC3 for Mega Dataset**

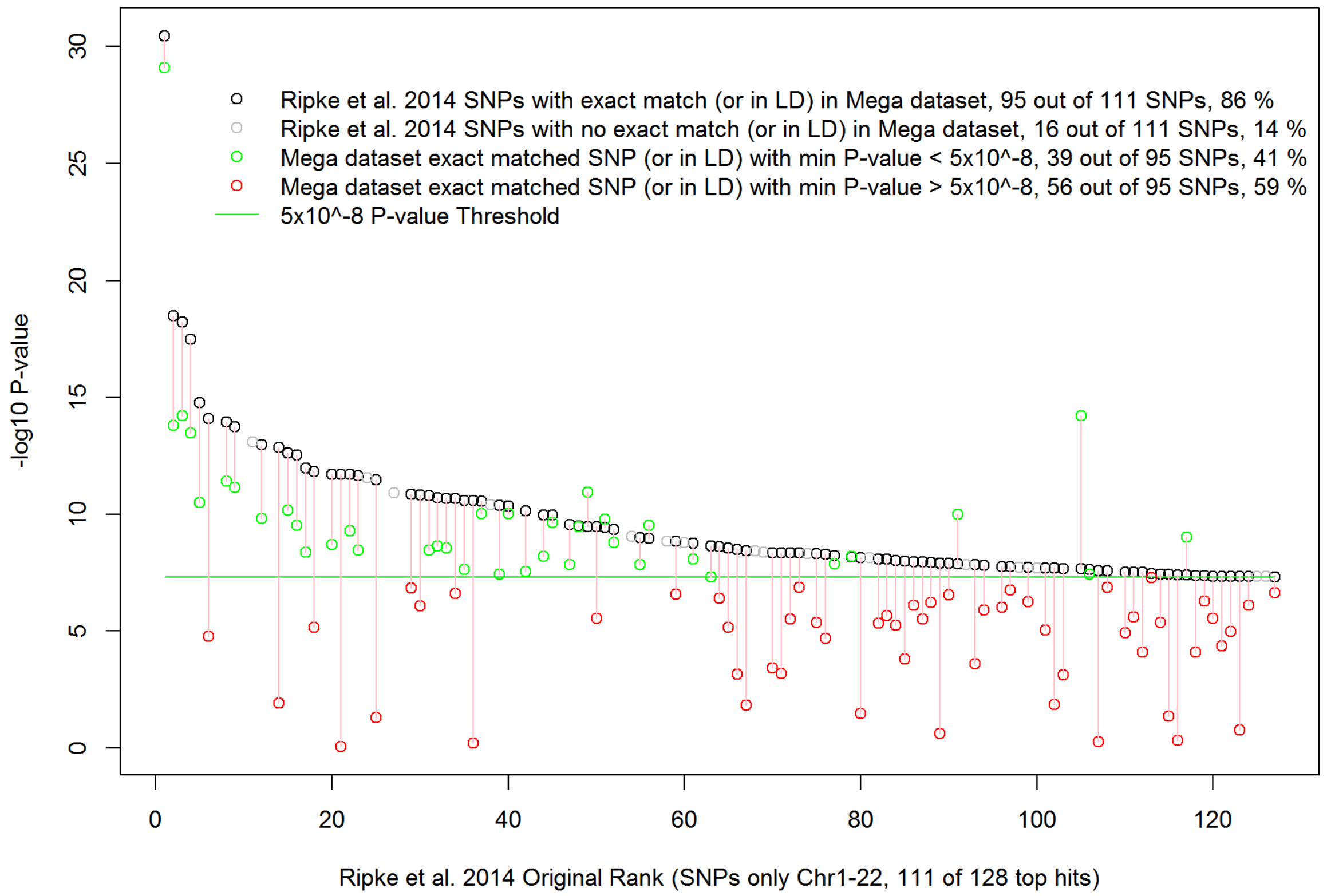

**Figure S3: Comparison with Ripke et al. 2014.** Comparison of Ripke et al 2014 [1] meta analysis results with mega analysis results here for the Mega dataset.

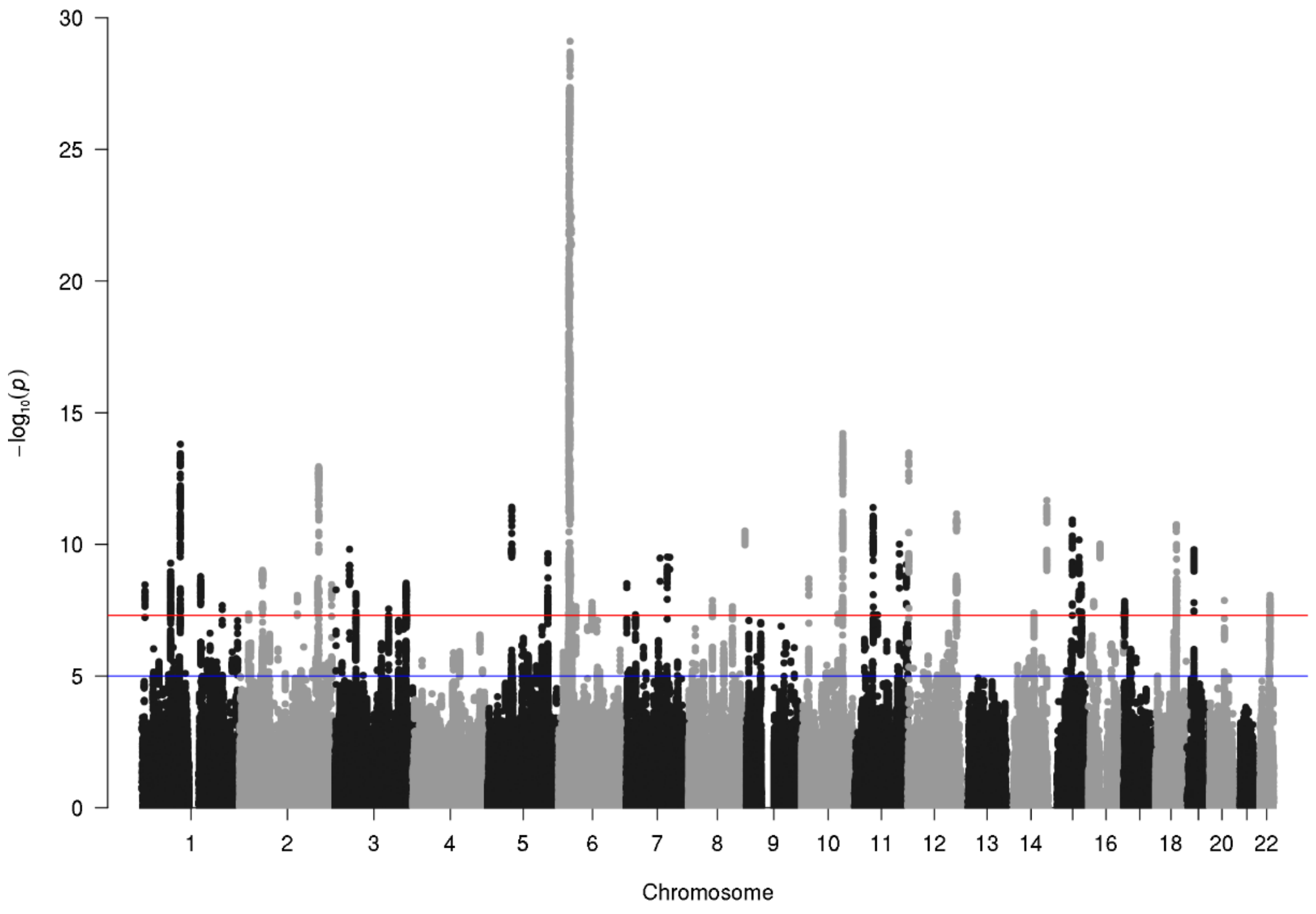

**Figure S4: Mega dataset association analysis results.**

## Computation

All calculations using the PGC data were carried out on the Lisa cluster which is part of the Dutch national e-infrastructure, where a repository of the data is maintained.

Computations were performed on the E5-2650 v2 nodes, each of which has 870Gb of scratch space, 64 Gb of memory, 20Mb of cache and clock speed 2.6 GHz. Each node supports 16 cores and up to 15 computations can be run simultaneously on each node. In practice, due to memory requirements at various points in the pipeline only one run at a time was carried out on a given node.

Approximate computation times on this hardware for the ANN (in serial mode) were as follows:

At the lower end of the analyses described (sample size approximately 10,000, number of SNPs approximately 1,300), run time was approximately 28 minutes, standard deviation 10 minutes as measured over 10 runs. At the upper end (sample size approximately 54,000, number of SNPs approximately 16,000) run time was approximately 3,530 minutes, standard deviation 620 minutes as measured over 10 runs.

Computation time was observed to vary approximately linearly with both sample size and number of SNPs over the range considered.

## Workflow

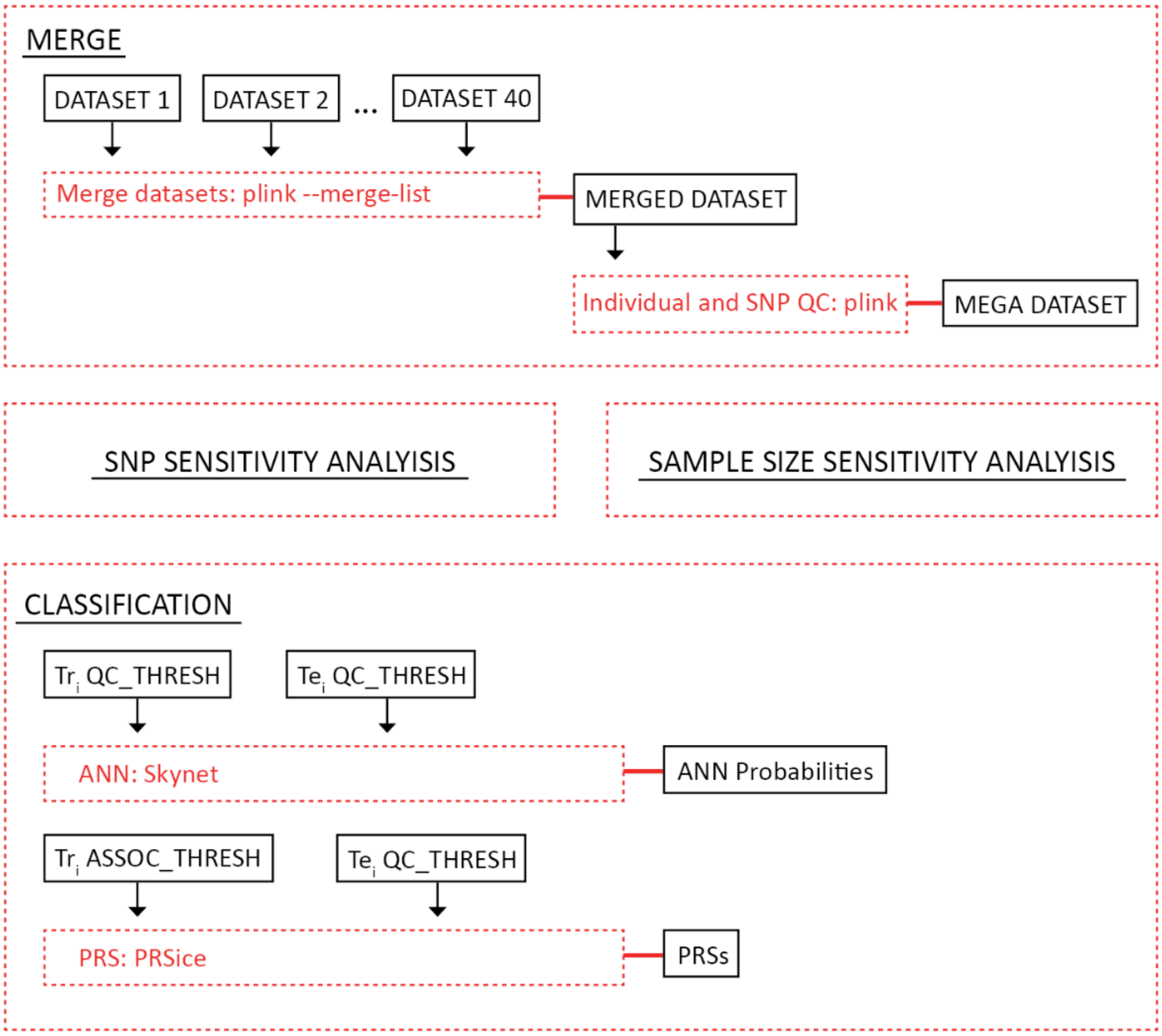

**Figure S5: Overall workflow for all analyses**

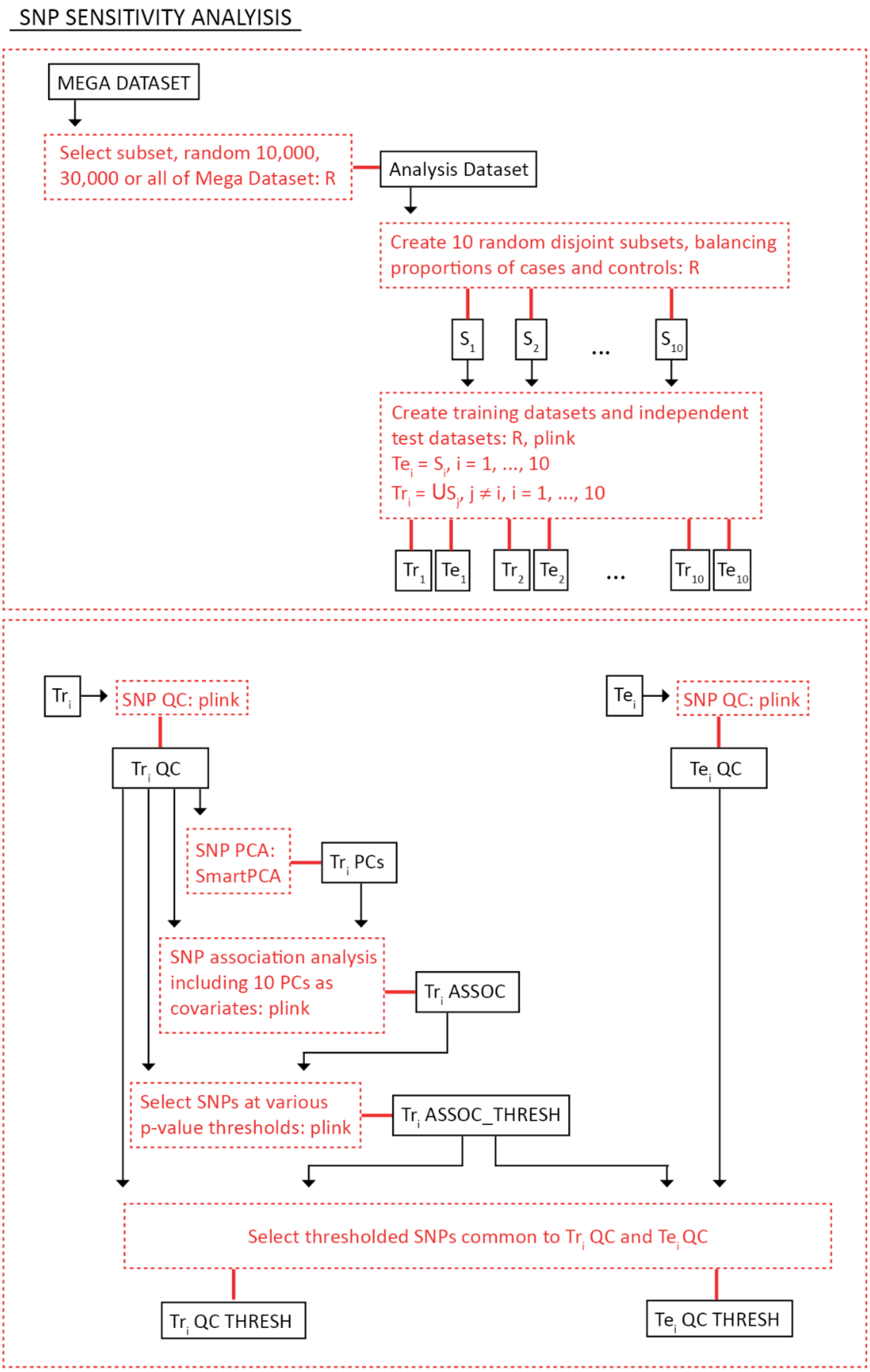

**Figure S6: Workflow for SNP sensitivity analysis**

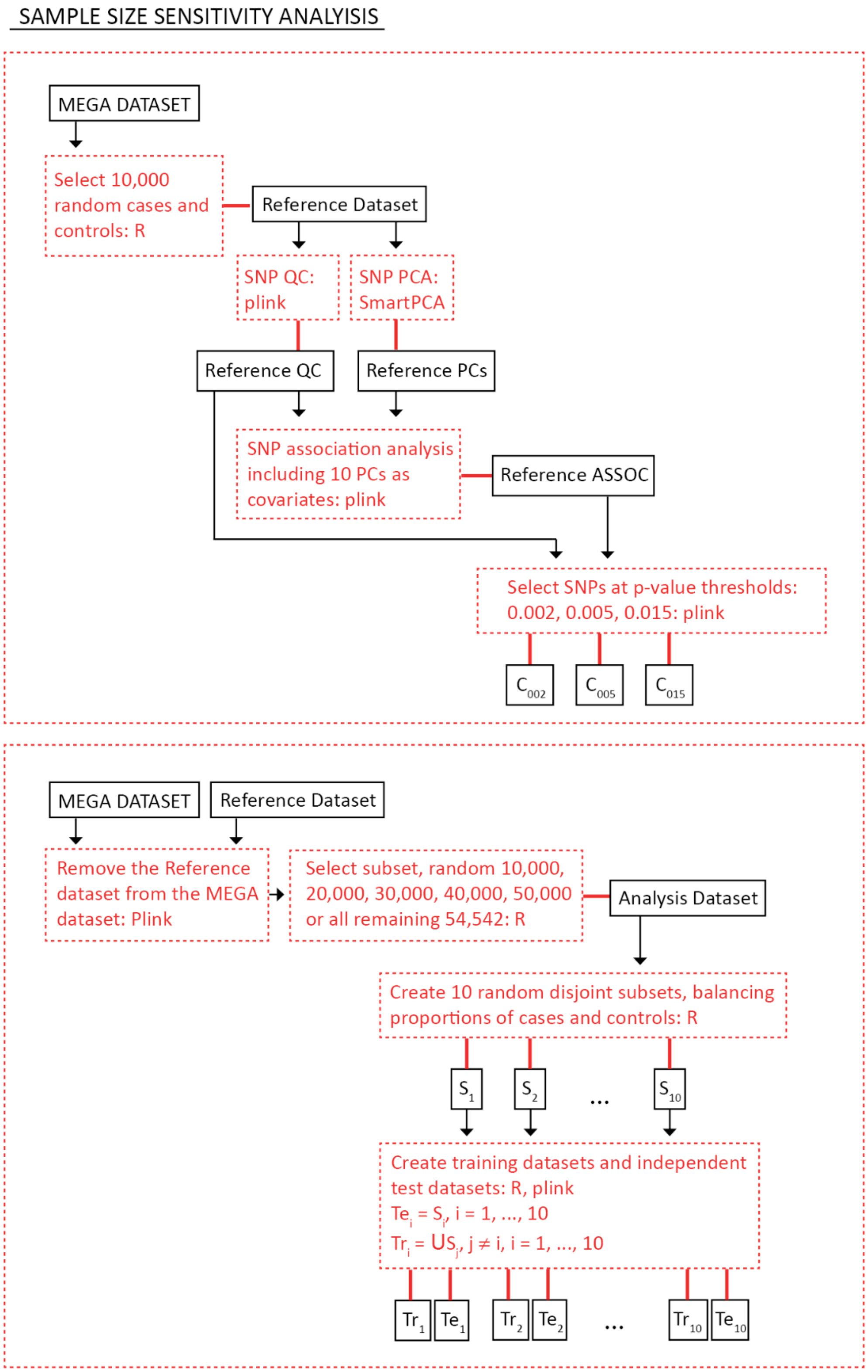

**Figure S7: Workflow for Sample Sensitivity Analysis: Creating SNP sets and train and test datasets to be prepared for the ANN and PRS**

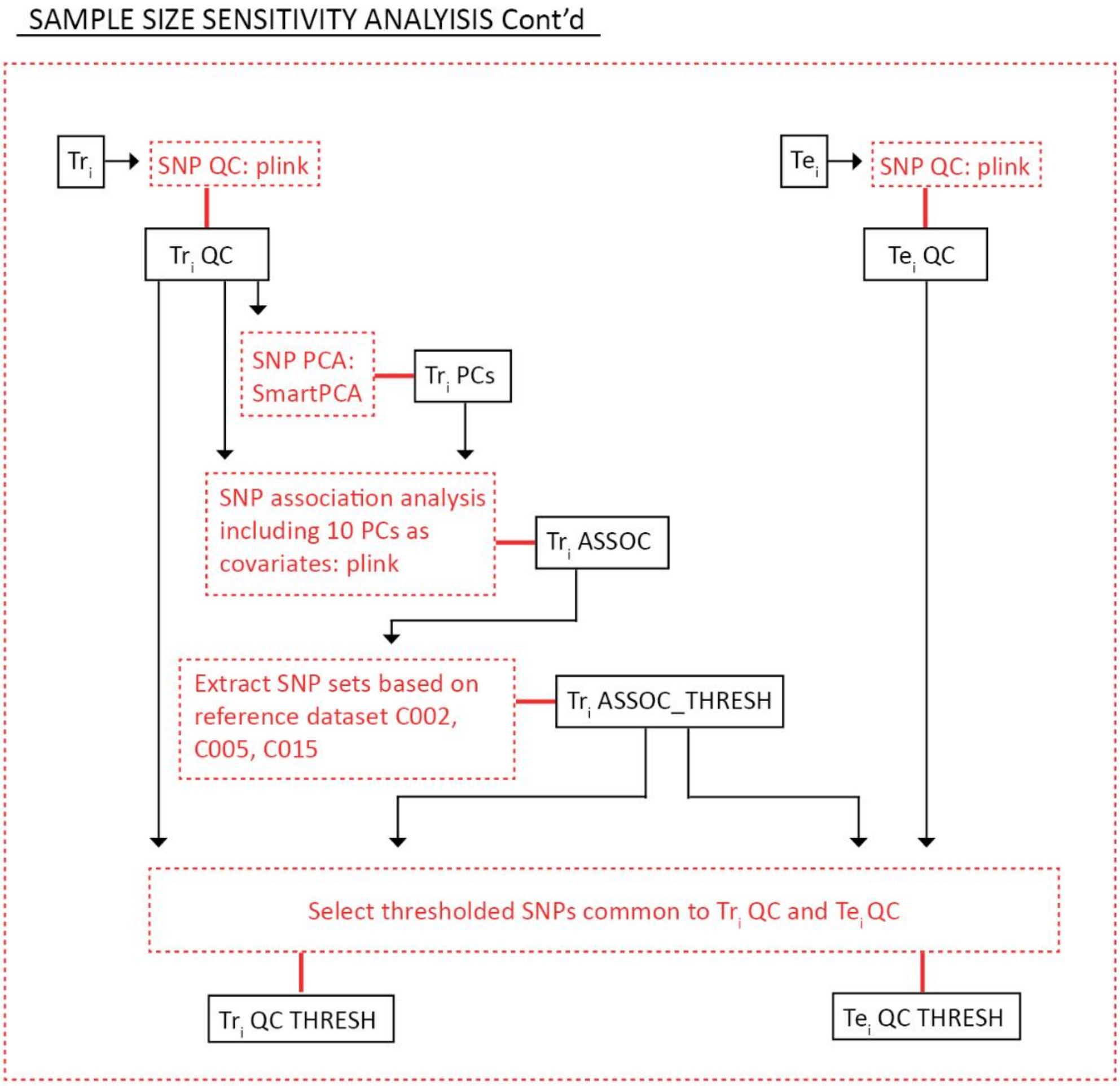

**Figure S8:** ***Workflow for Sample Sensitivity Analysis: Creating train and test datasets for the ANN and PRS***

## Schizophrenia Working Group of the Psychiatric Genomics Consortium

Stephan Ripke[1,2], Benjamin M. Neale[1,2,3,4], Aiden Corvin[5], James T. R. Walters[6], Kai-How Farh[1], Peter A. Holmans[6,7], Phil Lee[1,2,4], Brendan Bulik-Sullivan[1,2], David A.Collier[8,9], Hailiang Huang[1,3], Tune H. Pers[3,10,11], Ingrid Agartz[12,13,14], EsbenAgerbo[15,16,17], Margot Albus[18], Madeline Alexander[19], Farooq Amin[20,21], Silviu A.Bacanu[22], Martin Begemann[23], Richard A.

Belliveau Jr[2], Judit Bene[24,25], Sarah E.Bergen[2,26], Elizabeth Bevilacqua[2], Tim B. Bigdeli[22], Donald W. Black[27], Richard Bruggeman[28], Nancy G. Buccola[29], Randy L. Buckner [30,31,32], William Byerley[33],Wiepke Cahn[34],Guiqing Cai[35,36],Dominique Campion[37],Rita M.Cantor[38],Vaughan J.Carr[39,40], Noa Carrera[6], Stanley V. Catts[39,41], Kimberly D. Chambert[2], Raymond C. K.Chan[42], Ronald Y. L. Chen[43], Eric Y. H. Chen[43,44], Wei Cheng[45], Eric F. C. Cheung[46], Siow Ann Chong[47], C. Robert Cloninger[48], David Cohen[49], Nadine Cohen[50], PaulCormican[5], Nick Craddock[6,7], James J. Crowley[51], Michael Davidson[54], KennethL.Davis[36],Franziska Degenhardt[55,56], Jurgen Del Favero[57],Ditte Demontis[17,58,59], Dimitris Dikeos[60], Timothy Dinan[61], Srdjan Djurovic[14,62], GaryDonohoe[5,63], Elodie Drapeau[36], Jubao Duan[64,65], Frank Dudbridge[66], NaserDurmishi[67], Peter Eichhammer[68], Johan Eriksson[69,70,71], Valentina Escott-Price[6],Laurent Essioux[72], Ayman H. Fanous[73,74,75,76,] Martilias S. Farrell[51], Josef Frank[77],Lude Franke[78], Robert Freedman[79], Nelson B. Freimer[80], Marion Friedl[81], Joseph I.Friedman[36], Menachem Fromer[1,2,4,82], Giulio Genovese[2], Lyudmila Georgieva[6], InaGiegling[81,83], Paola Giusti-Rodríguez[51], Stephanie Godard[84], Jacqueline I.Goldstein[1,3], Vera Golimbet[85], Srihari Gopal[86], Jacob Gratten[87], Lieuwe de Haan[88], Christian Hammer[23], Marian L. Hamshere[6], Mark Hansen[89], Thomas Hansen[17,90], Vahram Haroutunian[36,91,92], Annette M. Hartmann[81], Frans A. Henskens[39,93,94],Stefan Herms[55,56,95], Joel N. Hirschhorn[3,11,96], Per Hoffmann[55,56,95], AndreaHofman[55,56], Mads V. Hollegaard[97], David M. Hougaard[97], Masashi Ikeda[98], IngeJoa[99], Antonio Julia`100, Rene´S. Kahn[34], Luba Kalaydjieva[101,102], SenaKarachanak-Yankova[103], Juha Karjalainen[78], David Kavanagh[6], Matthew C. Keller[104], James L. Kennedy[105,106,107], Andrey Khrunin[108], Yunjung Kim[51], Janis Klovins[109],James A. Knowles[110], Bettina Konte[81], Vaidutis Kucinskas[111], Zita AusreleKucinskiene[111], Hana Kuzelova-Ptackova[112], Anna K. Ka¨hler[26], ClaudineLaurent[19,113], Jimmy Lee Chee Keong[47,114], S. Hong Lee[87], Sophie E. Legge[6], Bernard Lerer[115], Miaoxin Li[43,44,116], Tao Li[117], Kung-Yee Liang[118], Jeffrey Lieberman[119],Svetlana Limborska[108], Carmel M. Loughland[39,120], Jan Lubinski[121], JoukoLönnqvist[122], Milan Macek Jr[112], Patrik K. E. Magnusson[26], Brion S. Maher[123],Wolfgang Maier[124], Jacques Mallet[125], Sara Marsal[100], Manuel Mattheisen[17,58,59,126],Morten Mattingsdal[14,127], Robert W. McCarley[128,129], Colm McDonald[130], Andrew M.McIntosh[131,132], Sandra Meier[77], Carin J. Meijer[88], Bela Melegh[24,25], IngridMelle[14,133], Raquelle I. Mesholam-Gately[128,134], Andres Metspalu[135], Patricia T.Michie[39,136], Lili Milani[135], Vihra Milanova[137], Younes Mokrab[8], Derek W. Morris[5,63],Ole Mors[17,58,138], Kieran C. Murphy[139], Robin M. Murray[140], Inez Myin-Germeys[141],Bertram Mu¨ller-Myhsok[142,143,144], Mari Nelis[135], Igor Nenadic[145], Deborah A. Nertney[146], Gerald Nestadt[147],Kristin K.Nicodemus[148], LieneNikitina-Zake[109], LauraNisenbaum[149], Annelie Nordin[150], Eadbhard O'Callaghan[151], Colm O'Dushlaine[2], F. Anthony O'Neill[152], Sang-Yun Oh[153], Ann Olincy[79], Line Olsen[17,90], Jim Van Os[141,154],Psychosis Endophenotypes International Consortium[155], Christos Pantelis[39,156], George N. Papadimitriou[60], Sergi Papiol[23], Elena Parkhomenko[36], Michele T. Pato[110],Tiina Paunio[157,158], Milica Pejovic-Milovancevic[159], Diana O. Perkins[160], OlliPietiläinen[158,161], Jonathan Pimm[53], Andrew J. Pocklington[6], John Powell[140], AlkesPrice[3,162], Ann E. Pulver[147], Shaun M. Purcell[82], Digby Quested[163], Henrik B.Rasmussen[17,90], Abraham Reichenberg[36], Mark A. Reimers[164], Alexander L. Richards[6], Joshua L. Roffman[30,32], Panos Roussos[82,165], Douglas M. Ruderfer[6,82],Veikko Salomaa[71], Alan R. Sanders[64,65], Ulrich Schall[39,120], Christian R. Schubert[166],Thomas G. Schulze[77,167], Sibylle G. Schwab[168], Edward M. Scolnick[2], Rodney J.Scott[39,169,170], Larry J. Seidman[128,134], Jianxin Shi1[71], Engilbert Sigurdsson[172],Teimuraz Silagadze[173], Jeremy M. Silverman[36,174], Kang Sim[47], Petr Slominsky[108],Jordan W. Smoller[2,4], Hon-Cheong So[43], Chris C. A. Spencer[175], Eli A. Stahl[3,82], HreinnStefansson[176], Stacy

Steinberg[176], Elisabeth Stogmann[177], Richard E. Straub[178], EricStrengman[179,34], Jana Strohmaier[77], T. Scott Stroup[119], Mythily Subramaniam[47], Jaana Suvisaari[122], Dragan M. Svrakic[48], Jin P. Szatkiewicz[51], Erik Sö̈derman[12], Srinivas Thirumalai[180], Draga Toncheva[103], Sarah Tosato[181], Juha Veijola[182,183],JohnWaddington[184], Dermot Walsh[185], Dai Wang[86], Qiang Wang[117], Bradley T. Webb[22],Mark Weiser[54], Dieter B. Wildenauer[186], Nigel M. Williams[6], Stephanie Williams[51], Stephanie H. Witt[77], Aaron R. Wolen[164], Emily H. M. Wong[43], Brandon K. Wormley[22],Hualin SimonXi[187], Clement C. Zai[105,106], Xuebin Zheng[188], FritzZimprich[177], Naomi R. Wray[87], Kari Stefansson[176], Peter M. Visscher[87], Wellcome Trust Case-Control Consortium 2[189], Rolf Adolfsson[150], Ole A. Andreassen[14,133], Douglas H. R.Blackwood[132], Elvira Bramon[190], Joseph D. Buxbaum[35,36,91,191], Anders D.Børglum[17,58,59,138], Sven Cichon[55,56,95,192], Ariel Darvasi[193], Enrico Domenici[194],Hannelore Ehrenreich[23],Tõnu Esko[3,11,96,135], Pablo V. Gejman[64,65], Michael Gill[5],Hugh Gurling[53], Christina M. Hultman[26], Nakao Iwata[98], Assen V.Jablensky[39,102,186,195], Erik G. Jö̈nsson[12,14], Kenneth S. Kendler[196], George Kirov[6], Jo Knight[105,106,107],Todd Lencz[197,198,199], Douglas F. Levinson[19],Qingqin S. Li[86],Jianjun Liu[188,200], Anil K. Malhotra[197,198,199], Steven A. McCarroll[2,96], Andrew McQuillin[53], Jennifer L. Moran[2], Preben B. Mortensen[15,16,17], Bryan J. Mowry[87,201], Markus M. Nöthen[55,56], Roel A. Ophoff[38,80,34], Michael J. Owen[6,7], Aarno Palotie[2,4,161], Carlos N. Pato[110], Tracey L. Petryshen[2,128,202], Danielle Posthuma[203,204,205], Marcella Rietschel[77], Brien P. Riley[196], Dan Rujescu[81,83], Pak C. Sham[43,44,116], Pamela Sklar[82,91,165], David St Clair[206], Daniel R. Weinberger[178,207], Jens R. Wendland[166],Thomas Werge[17,90,208], Mark J. Daly[1,2,3], Patrick F. Sullivan[26,51,160] & Michael C.O'Donovan[6,7]

[1]Analytic and Translational Genetics Unit, Massachusetts General Hospital, Boston, Massachusetts 02114, USA. [2]Stanley Center for Psychiatric Research, Broad Institute of MIT and Harvard, Cambridge, Massachusetts 02142, USA. [3]Medical and Population Genetics Program, Broad Institute of MIT and Harvard, Cambridge, Massachusetts 02142, USA.[4]Psychiatric and Neurodevelopmental Genetics Unit, Massachusetts General Hospital, Boston, Massachusetts 02114, USA.[5]Neuropsychiatric Genetics Research Group, Department of Psychiatry, Trinity College Dublin, Dublin 8, Ireland.[6]MRC Centre for Neuropsychiatric Genetics and Genomics, Institute of Psychological Medicine and Clinical Neurosciences, School of Medicine, Cardiff University,CardiffCF244HQ, UK. [7]NationalCentreforMentalHealth,CardiffUniversity,Cardiff CF244HQ, UK.[8]EliLilly and Company Limited, Erl Wood Manor, Sunninghill Road, Windlesham, Surrey GU20 6PH, UK.[9]Social, Genetic and Developmental Psychiatry Centre, Institute of Psychiatry, King's College London,LondonSE58AF,UK.[10]Centerfor Biological Sequence Analysis, Department of Systems Biology, Technical University of Denmark, DK-2800,Denmark.[11]Division of Endocrinology and Center for Basic and Translational Obesity Research, Boston Children's Hospital, Boston, Massachusetts 02115,USA.[12]Departmentof Clinical Neuroscience, Psychiatry Section, Karolinska Institutet, SE-17176 Stockholm, Sweden.[13]Department of Psychiatry, Diakonhjemmet Hospital, 0319 Oslo, Norway.[14]NORMENT, KG Jebsen Centre for Psychosis Research, Institute of Clinical Medicine, University of Oslo, 0424 Oslo, Norway.[15]Centre for Integrative Register-based Research, CIRRAU, Aarhus University, DK-8210 Aarhus, Denmark.[16]National Centre for Register-based Research, Aarhus University,DK-8210 Aarhus,Denmark.[17]TheLundbeckFoundation Initiative for Integrative Psychiatric Research, iPSYCH, Denmark.[18]StateMental Hospital, 85540 Haar, Germany.[19]Department of Psychiatry and Behavioral Sciences, Stanford University, Stanford, California 94305,

USA.[20]Department of Psychiatry and Behavioral Sciences, Atlanta Veterans Affairs Medical Center, Atlanta, Georgia 30033, USA.[21]Department of Psychiatry and Behavioral Sciences, Emory University, Atlanta, Georgia 30322, USA. [22]Virginia Institute for Psychiatric and Behavioral Genetics, Department of Psychiatry, Virginia Commonwealth University, Richmond, Virginia 23298, USA.[23]Clinical Neuroscience, Max Planck Institute of Experimental Medicine, Göttingen 37075, Germany.[24]Department of Medical Genetics, University of Pe´cs, Pe´cs H-7624, Hungary.[25]Szentagothai Research Center, University of Pécs, Pécs H-7624, Hungary. [26]Department of Medical Epidemiology and Biostatistics, Karolinska Institutet, Stockholm SE-17177, Sweden.[27]Department of Psychiatry, University of Iowa Carver College of Medicine, Iowa City, Iowa 52242, USA.[28]University Medical Center Groningen, Department of Psychiatry, University of Groningen NL-9700 RB, The Netherlands.[29]School of Nursing, Louisiana State University Health Sciences Center, New Orleans, Louisiana 70112, USA.[30]Athinoula A. Martinos Center, Massachusetts General Hospital, Boston, Massachusetts 02129, USA.[31]Center for Brain Science, Harvard University, Cambridge, Massachusetts 02138, USA.[32]Department of Psychiatry, Massachusetts General Hospital, Boston, Massachusetts 02114, USA.[33]Department of Psychiatry, University of California at San Francisco, San Francisco, California 94143,USA.[34]University Medical Center Utrecht, Department of Psychiatry, Rudolf Magnus Institute of Neuroscience, 3584 Utrecht, The Netherlands.[35]Department of Human Genetics, Icahn School of Medicine at Mount Sinai, New York, New York 10029, USA.[36]Department of Psychiatry, Icahn School of Medicine at Mount Sinai, New York, New York 10029, USA.[37]Centre Hospitalier du Rouvray and INSERM U1079 Faculty of Medicine,76301 Rouen, France.[38]Department of Human Genetics, David Geffen School of Medicine, University of California, Los Angeles, California 90095, USA.[39]Schizophrenia Research Institute,  Sydney NSW2010, Australia.[40]School of Psychiatry, University of New South Wales, Sydney NSW 2031, Australia.[41]Royal Brisbane and Women's Hospital, University of Queensland, Brisbane, St Lucia QLD 4072, Australia.[42]Institute of Psychology, Chinese Academy of Science, Beijing 100101, China.[43]Department of Psychiatry, Li Ka Shing Faculty of Medicine, The University of Hong Kong, Hong Kong,China.[44]State Key Laboratory for Brain and Cognitive Sciences, Li Ka Shing Faculty of Medicine, The University of Hong Kong, Hong Kong, China.[45]Department of Computer Science, University of North Carolina, Chapel Hill, North Carolina 27514, USA.[46]CastlePeak Hospital, Hong Kong, China.[47]Institute of Mental Health, Singapore 539747, Singapore.[48]Department of Psychiatry, Washington University, St. Louis, Missouri 63110, USA.[49]DepartmentofChild and Adolescent Psychiatry, Assistance Publique Hopitaux de Paris, Pierre and Marie Curie Faculty of Medicine and Institute for Intelligent Systems and Robotics, Paris 75013, France.[50]Blue Note Biosciences, Princeton, New Jersey 08540,USA.[51]Department of Genetics, University of North Carolina, Chapel Hill, North Carolina 27599-7264, USA. [53]Molecular Psychiatry Laboratory, Division of Psychiatry, University College London, London WC1E6JJ, UK.[54]Sheba Medical Center, Tel Hashomer 52621, Israel.[55]Department of Genomics, Life and Brain Center, D-53127 Bonn,Germany.[56]Institute of Human Genetics, University of Bonn, D-53127 Bonn, Germany.[57]Applied Molecular Genomics Unit, VIB Department of Molecular Genetics, University of Antwerp, B-2610 Antwerp, Belgium.[58]Centre for Integrative Sequencing, iSEQ, Aarhus University, DK-8000 Aarhus C, Denmark.[59]Department of Biomedicine, Aarhus University, DK-8000 Aarhus C, Denmark.[60]First Department of Psychiatry, University of Athens Medical School, Athens 11528, Greece.[61]Department of Psychiatry, University College Cork, Co. Cork, Ireland.62Department of Medical Genetics, Oslo University Hospital, 0424 Oslo, Norway.[63]Cognitive Genetics and Therapy Group, School of Psychology and Discipline of Biochemistry, National University of Ireland Galway, Co. Galway, Ireland.[64]Department of Psychiatry and Behavioral Neuroscience, University of Chicago, Chicago, Illinois 60637, USA.[65]Department of Psychiatry and Behavioral Sciences, NorthShore University Health System, Evanston, Illinois 60201, USA.[66]Department of Non-Communicable Disease Epidemiology, London School of Hygiene and Tropical Medicine, London WC1E 7HT, UK.[67]Department of Child and Adolescent Psychiatry, University Clinic of Psychiatry, Skopje 1000, Republic of Macedonia.[68]Department of Psychiatry, University of Regensburg, 93053 Regensburg, Germany.[69]Department of General Practice, Helsinki University Central Hospital, University of Helsinki P.O. Box 20, Tukholmankatu 8 B, FI-00014, Helsinki, Finland.[70]Folkhälsan Research Center, Helsinki, Finland, Biomedicum Helsinki 1, Haartmaninkatu 8, FI-00290,Helsinki, Finland.[71]National Institute for Health and Welfare, P.O. Box 30, FI-00271Helsinki, Finland.[72]Translational Technologies and Bioinformatics, Pharma Research and Early Development, F. Hoffman-La Roche, CH-4070 Basel, Switzerland.[73]Department of Psychiatry, Georgetown University School of Medicine, Washington DC20057, USA.[74]Department of Psychiatry, Keck School of Medicine of the University of Southern California, Los Angeles, California 90033, USA.[75]Department of Psychiatry, Virginia Commonwealth University School of Medicine, Richmond, Virginia 23298, USA.[76]Mental Health Service Line, Washington VA Medical Center, Washington DC 20422,USA.[77]Department of Genetic Epidemiology in Psychiatry, Central Institute of Mental Health, Medical Faculty Mannheim, University of Heidelberg, Heidelberg , D-68159Mannheim, Germany.[78]Department of Genetics, University of Groningen, University Medical Centre Groningen, 9700 RB Groningen, The Netherlands.[79]Department of Psychiatry, University of Colorado Denver, Aurora, Colorado 80045, USA.[80]Center for Neurobehavioral Genetics, Semel Institute for Neuroscience and Human Behavior, University of California, Los Angeles, California 90095, USA.[81]Department of Psychiatry, University of Halle, 06112 Halle, Germany.[82]Division of Psychiatric Genomics, Department of Psychiatry, Icahn School of Medicine at Mount Sinai, New York, New York, New York 10029, USA.[83]Department of Psychiatry, University of Munich, 80336, Munich, Germany. [84]Departments of Psychiatry and Human and Molecular Genetics, INSERM, Institut de Myologie, Hôpital de la Pitiè-Salpêtrière, Paris 75013, France.[85]Mental Health Research Centre, Russian Academy of Medical Sciences, 115522 Moscow, Russia.[86]Neuroscience Therapeutic Area, Janssen Research and Development, Raritan, New Jersey 08869, USA.[87]Queensland Brain Institute, The University of Queensland, Brisbane, Queensland, QLD 4072, Australia.[88]Academic Medical Centre University of Amsterdam, Department of Psychiatry, 1105 AZ Amsterdam, The Netherlands.[89]Illumina, La Jolla, California, California 92122, USA.[90]Institute of Biological Psychiatry, Mental Health Centre Sct. Hans, Mental Health Services Copenhagen,DK-4000,Denmark.[91]FriedmanBrain Institute, Icahn School of Medicine at Mount Sinai, New York, New York10029, USA.[92]J. J. Peters VA Medical Center, Bronx, New York, New York 10468, USA.[93]PriorityResearchCentrefor Health Behaviour, University of Newcastle, NewcastleNSW2308, Australia.[94]School of Electrical Engineering and Computer Science, University of Newcastle, Newcastle NSW 2308, Australia.[95]Division of Medical Genetics, Department of Biomedicine, University of Basel, Basel CH-4058, Switzerland.[96]Department of Genetics, Harvard Medical School, Boston, Massachusetts, Massachusetts 02115, USA.[97]Section of Neonatal Screening and Hormones, Department of Clinical Biochemistry, Immunology and Genetics, Statens Serum Institut, Copenhagen DK-2300, Denmark.[98]Department of Psychiatry, Fujita Health University School of Medicine, Toyoake, Aichi,470-1192, Japan.[99]Regional Centre for Clinical Research in Psychosis,

Department of Psychiatry, Stavanger University Hospital, 4011 Stavanger, Norway. [100]Rheumatology Research Group, Vall d'Hebron Research Institute, Barcelona 08035, Spain.[101]Centre for Medical Research, The University of Western Australia, Perth WA6009, Australia.[102]ThePerkins Institute for Medical Research, The University of Western Australia, PerthWA6009, Australia.[103]Department of Medical Genetics, Medical University, Sofia 1431,Bulgaria.[104]Department of Psychology, University of Colorado Boulder, Boulder, Colorado 80309, USA.[105]Campbell Family Mental Health Research Institute, Centre for Addiction and Mental Health, Toronto, Ontario M5T 1R8, Canada.[106]Department of Psychiatry, University of Toronto, Toronto, Ontario M5T 1R8, Canada.[107]Institute of Medical Science, University of Toronto, Toronto, Ontario M5S1A8, Canada.[108]Institute of Molecular Genetics, Russian Academy of Sciences, Moscow 123182, Russia.[109]Latvian Biomedical Research and Study Centre, Riga, LV-1067, Latvia.[110]Department of Psychiatry and Zilkha Neurogenetics Institute, Keck School of Medicine at University of Southern California, Los Angeles, California 90089, USA.[111]Faculty of Medicine, Vilnius University, LT-01513 Vilnius, Lithuania.[112]Department of Biology and Medical Genetics,2nd Faculty of Medicine and University Hospital Motol, 150 06 Prague, Czech Republic.[113]Department of Child and Adolescent Psychiatry, Pierre and Marie Curie Faculty of Medicine, Paris 75013, France.[114]Duke-NUS Graduate Medical School, Singapore169857.[115]Department of Psychiatry, Hadassah-Hebrew University Medical Center, Jerusalem 91120, Israel.[116]Centre for Genomic Sciences, The University of Hong Kong, Hong Kong, China.[117]Mental Health Centre and Psychiatric Laboratory, West China Hospital, Sichuan University, Chengdu, 610041 Sichuan, China.[118]Department of Biostatistics, Johns Hopkins University Bloomberg School of Public Health, Baltimore, Maryland 21205, USA.[119]Department of Psychiatry, Columbia University, New York, New York 10032, USA.[120]Priority Centre for Translational Neuroscience and Mental Health, University of Newcastle, Newcastle NSW 2300, Australia.[121]Department of Genetics and Pathology, International Hereditary Cancer Center, Pomeranian Medical University in Szczecin, 70-453 Szczecin, Poland.[122]Department of Mental Health and Substance Abuse Services; National Institute for Health and Welfare, P.O.BOX30, FI-00271 Helsinki, Finland.[123]Department of Mental Health, Bloomberg School of Public Health, Johns Hopkins University, Baltimore, Maryland 21205, USA.[124]Department of Psychiatry, University of Bonn, D-53127 Bonn, Germany.[125]Centre National de la Recherche Scientifique, Laboratoire de Génétique Moléculaire de la Neurotransmission et des Processus Neurodégénératifs, Hôpital de la Pitié Salpêtrière, 75013 Paris, France.[126]Department of Genomics Mathematics, University of Bonn, D-53127 Bonn, Germany.[127]Research Unit, Sørlandet Hospital, 4604 Kristiansand, Norway.[128]Department of Psychiatry, Harvard Medical School, Boston, Massachusetts 02115, USA.[129]VA BostonHealthCareSystem,Brockton,Massachusetts02301,USA.[130]Department of Psychiatry, National University of Ireland Galway, Co. Galway, Ireland.[131]Centre for Cognitive Ageing and Cognitive Epidemiology, University of Edinburgh, Edinburgh EH16 4SB, UK.[132]Division of Psychiatry, University of Edinburgh, Edinburgh EH16 4SB, UK.[133]Divisionof Mental Health and Addiction, Oslo University Hospital, 0424 Oslo, Norway.[134]Massachusetts Mental Health Center Public Psychiatry Division of the Beth Israel Deaconess Medical Center, Boston, Massachusetts 02114, USA.[135]Estonian Genome Center, University of Tartu, Tartu 50090, Estonia.[136]School of Psychology, University of Newcastle, Newcastle NSW2308, Australia.[137]First Psychiatric Clinic, Medical University, Sofia 1431, Bulgaria.[138]Department P, Aarhus University Hospital, DK-8240 Risskov,Denmark.[139]Department of Psychiatry, Royal College of Surgeons in Ireland, Dublin 2,Ireland.[140]King's College London, London SE58AF,

UK.[141]Maastricht University Medical Centre, South Limburg Mental Health Research and Teaching Network,EURON,6229HX Maastricht, The Netherlands.[142]Institute of Translational Medicine, University of Liverpool, Liverpool L69 3BX, UK.[143]Max Planck Institute of Psychiatry, 80336 Munich, Germany.[144]Munich Cluster for Systems Neurology (SyNergy), 80336 Munich,Germany.[145]Department of Psychiatry and Psychotherapy, Jena University Hospital, 07743 Jena,Germany.[146]Department of Psychiatry, Queensland Brain Institute and Queensland Centre for Mental Health Research, University of Queensland, Brisbane, Queensland, St Lucia QLD 4072, Australia.147Department of Psychiatry and Behavioral Sciences, Johns Hopkins University School of Medicine, Baltimore, Maryland 21205, USA.[148]Department of Psychiatry, Trinity College Dublin, Dublin 2, Ireland.[149]Eli Lilly and Company, Lilly Corporate Center, Indianapolis, 46285 Indiana, USA.[150]Department of Clinical Sciences, Psychiatry, Umea˚ University, SE-901 87 Umea˚, Sweden.[151]DETECT Early Intervention Service for Psychosis, Blackrock, Co. Dublin, Ireland.[152]Centre for Public Health, Institute of Clinical Sciences, Queen's University Belfast, Belfast BT12 6AB, UK.[153] Lawrence Berkeley National Laboratory, University of California at Berkeley, Berkeley, California 94720, USA.[154]Institute of Psychiatry, King's College London, London SE5 8AF, UK. [156]Melbourne Neuropsychiatry Centre, University of Melbourne & Melbourne Health, Melbourne, Vic3053, Australia.[157]Department of Psychiatry, University of Helsinki, P.O. Box 590,FI-00029 HUS, Helsinki, Finland.[158]Public Health Genomics Unit, National Institute for Health and Welfare, P.O. BOX 30, FI-00271 Helsinki, Finland.[159]Medical Faculty, University of Belgrade, 11000 Belgrade, Serbia.[160]Department of Psychiatry, University of North Carolina, Chapel Hill, North Carolina 27599-7160, USA.[161]Institute for Molecular Medicine Finland, FIMM, University of Helsinki, P.O. Box 20FI-00014, Helsinki,Finland.[162]Department of Epidemiology, Harvard School of Public Health, Boston, Massachusetts 02115, USA.[163]Department of Psychiatry, University of Oxford, Oxford,OX3 7JX, UK.[164]Virginia Institute for Psychiatric and Behavioral Genetics, Virginia Commonwealth University, Richmond, Virginia 23298, USA.[165]Institute for Multiscale Biology, Icahn School of Medicine at Mount Sinai, New York, New York 10029, USA.[166]Pharma Therapeutics Clinical Research, Pfizer Worldwide Research and Development, Cambridge, Massachusetts 02139, USA.[167]Department of Psychiatry and Psychotherapy, University of Gottingen, 37073 Göttingen, Germany.[168]Psychiatry and Psychotherapy Clinic, University of Erlangen, 91054 Erlangen, Germany.[169]Hunter New England Health Service, Newcastle NSW 2308, Australia.[170]School of Biomedical Sciences, University of Newcastle, Newcastle NSW 2308, Australia.[171]Division of Cancer Epidemiology and Genetics, National Cancer Institute, Bethesda, Maryland 20892, USA.[172]University of Iceland, Landspitali, National University Hospital, 101 Reykjavik, Iceland.[173]Department of Psychiatry and Drug Addiction, Tbilisi State Medical University (TSMU), N33, 0177 Tbilisi, Georgia.[174]Research and Development, Bronx Veterans Affairs Medical Center, New York, New York 10468, USA.[175]Wellcome Trust Centre for Human Genetics, Oxford OX3 7BN, UK.[176]deCODE Genetics, 101 Reykjavik, Iceland.[177]Department of Clinical Neurology, Medical University of Vienna, 1090 Wien, Austria.[178]Lieber Institute for Brain Development, Baltimore, Maryland 21205, USA.[179]Department of Medical Genetics, University Medical Centre Utrecht, Universiteitsweg 100, 3584CG, Utrecht, The Netherlands.[180]Berkshire Healthcare NHS Foundation Trust, Bracknell RG12 1BQ, UK. [181]Section of Psychiatry, University of Verona, 37134 Verona, Italy.[182]Department of Psychiatry, University of Oulu, P.O. Box 5000, 90014, Finland.[183]University Hospital of Oulu, P.O. Box 20, 90029 OYS, Finland.[184]Molecular and Cellular Therapeutics, Royal College of Surgeons in Ireland, Dublin 2, Ireland.[185]Health

Research Board, Dublin 2,Ireland.[186]School of Psychiatry and Clinical Neurosciences, The University of Western Australia, Perth WA6009, Australia.[187]Computational Sciences CoE, Pfizer Worldwide Research and Development, Cambridge, Massachusetts 02139, USA.[188]Human Genetics, Genome Institute of Singapore, A*STAR, Singapore 138672. [190]University College London, London WC1E 6BT, UK.[191]Department of Neuroscience, Icahn School of Medicine at Mount Sinai, New York, New York 10029, USA.[192]Institute of Neuroscience and Medicine(INM-1), Research Center Juelich, 52428 Juelich, Germany.[193]Department of Genetics, The Hebrew University of Jerusalem, 91905 Jerusalem, Israel.[194]Neuroscience Discovery and Translational Area, Pharma Research and Early Development, F. Hoffman-La Roche,CH-4070 Basel, Switzerland.[195]Centre for Clinical Research in Neuropsychiatry, School of Psychiatry and Clinical Neurosciences, The University of Western Australia, Medical Research Foundation Building, Perth WA6000, Australia.[196]Virginia Institute for Psychiatric and Behavioral Genetics, Departments of Psychiatry and Human and Molecular Genetics, Virginia Commonwealth University, Richmond, Virginia 23298, USA.[197]The Feinstein Institute for Medical Research, Manhasset, New York 11030, USA.[198]The Hofstra NS-LIJ School of Medicine, Hempstead, New York 11549, USA.[199]The Zucker Hillside Hospital, Glen Oaks, New York 11004, USA.[200]Saw Swee Hock School of Public Health, National University of Singapore, Singapore 117597, Singapore.[201]Queensland Centre for Mental Health Research, University of Queensland, Brisbane4076, Queensland, Australia.[202]Center for Human Genetic Research and Department of Psychiatry, Massachusetts General Hospital, Boston, Massachusetts 02114, USA.[203]Department of Child and Adolescent Psychiatry, Erasmus University Medical Centre, Rotterdam 3000, The Netherlands.[204]Department of Complex Trait Genetics, Neuroscience Campus Amsterdam, VU University Medical Center Amsterdam, Amsterdam 1081, The Netherlands.[205]Department of Functional Genomics, Center for Neurogenomics and Cognitive Research, Neuroscience Campus Amsterdam, VU University, Amsterdam 1081, The Netherlands.[206]University of Aberdeen, Institute of Medical Sciences, Aberdeen AB25 2ZD, UK.[207]Departments of Psychiatry, Neurology, Neuroscience and Institute of Genetic Medicine, Johns Hopkins School of Medicine, Baltimore, Maryland 21205, USA.[208]Department of Clinical Medicine, University of Copenhagen, Copenhagen 2200, Denmark.

[155]Psychosis Endophenotype International Consortium
Maria J Arranz [156,234], Steven Bakker [101], Stephan Bender 235,236, Elvira Bramon [156,237,238], David A Collier [8,9], Benedicto Crespo-Facorro [239,240], Jeremy Hall [134], Conrad Iyegbe [156], Assen V Jablensky [241], René S Kahn [101], Luba Kalaydjieva [102,242], Stephen Lawrie [134], Cathryn M Lewis [156], Kuang Lin [156], Don H Linszen [243], Ignacio Mata [239,240], Andrew M McIntosh [134], Robin M Murray [142], Roel A Ophoff [80], Jim Van Os [143,156], John Powell [156], Dan Rujescu [81,83], Muriel Walshe [156], Matthias Weisbrod [236], Durk Wiersma [244].

[189]Wellcome Trust Case-Control Consortium 2
Management Committee: Peter Donnelly [180,217], Ines Barroso [218], Jenefer M Blackwell [219,220], Elvira Bramon [196], Matthew A Brown [221], Juan P Casas [222,223], Aiden Corvin [5], Panos Deloukas [218], Audrey Duncanson [224], Janusz Jankowski [225], Hugh S Markus [226], Christopher G Mathew [227], Colin N A Palmer [228], Robert Plomin [9], Anna Rautanen [180], Stephen J Sawcer [229], Richard C Trembath [227], Ananth C Viswanathan [230,231], Nicholas W Wood [232]. Data and Analysis Group: Chris C A Spencer [180], Gavin Band [180], Céline Bellenguez [180], Peter Donnelly [180,217], Colin

Freeman [180], Eleni Giannoulatou [180], Garrett Hellenthal [180], Richard Pearson [180], Matti Pirinen [180], Amy Strange [180], Zhan Su [180], Damjan Vukcevic [180]. DNA, Genotyping, Data QC, and Informatics: Cordelia Langford [218], Ines Barroso [218], Hannah Blackburn [218], Suzannah J Bumpstead [218], Panos Deloukas [218], Serge Dronov [218], Sarah Edkins [218], Matthew Gillman [218], Emma Gray [218], Rhian Gwilliam [218], Naomi Hammond [218], Sarah E Hunt [218], Alagurevathi Jayakumar [218], Jennifer Liddle [218], Owen T McCann [218], Simon C Potter [218], Radhi Ravindrarajah [218], Michelle Ricketts [218], Avazeh Tashakkori-Ghanbaria [218], Matthew Waller [218], Paul Weston [218], Pamela Whittaker [218], Sara Widaa [218]. Publications Committee: Christopher G Mathew [227], Jenefer M Blackwell [219,220], Matthew A Brown [221], Aiden Corvin [5], Mark I McCarthy [233], Chris C A Spencer [180].

[217] Department of Statistics, University of Oxford, Oxford, UK. [218] Wellcome Trust Sanger Institute, Wellcome Trust Genome Campus, Hinxton, Cambridge, UK. [219] Cambridge Institute for Medical Research, University of Cambridge School of Clinical Medicine, Cambridge, UK. [220] Telethon Institute for Child Health Research, Centre for Child Health Research, University of Western Australia, Subiaco, Western Australia, Australia. [221] Diamantina Institute of Cancer, Immunology and Metabolic Medicine, Princess Alexandra Hospital, University of Queensland, Brisbane, Queensland, Australia. [222] Department of Epidemiology and Population Health, London School of Hygiene and Tropical Medicine, London, UK. [223] Department of Epidemiology and Public Health, University College London, London, UK. [224] Molecular and Physiological Sciences, The Wellcome Trust, London, UK. [225] Peninsula School of Medicine and Dentistry, Plymouth University, Plymouth, UK. [226] Clinical Neurosciences, St George's University of London, London, UK. [227] Department of Medical and Molecular Genetics, School of Medicine, King's College London, Guy's Hospital, London, UK. [228] Biomedical Research Centre, Ninewells Hospital and Medical School, Dundee, UK. [229] Department of Clinical Neurosciences, University of Cambridge, Addenbrooke's Hospital, Cambridge, UK. [230] Institute of Ophthalmology, University College London, London, UK. [231] National Institute for Health Research, Biomedical Research Centre at Moorfields Eye Hospital, National Health Service Foundation Trust, London, UK. [232] Department of Molecular Neuroscience, Institute of Neurology, London, UK. [233] Oxford Centre for Diabetes, Endocrinology and Metabolism, Churchill Hospital, Oxford, UK. [234] Fundació de Docència i Recerca Mútua de Terrassa, Universitat de Barcelona, Spain. [235] Child and Adolescent Psychiatry, University of Technology Dresden, Dresden, Germany. [236] Section for Experimental Psychopathology, General Psychiatry, Heidelberg, Germany. [237] Institute of Cognitive Neuroscience, University College London, London, UK. [238] Mental Health Sciences Unit, University College London, London, UK. [239] Centro Investigación Biomédica en Red Salud Mental, Madrid, Spain. [240] University Hospital Marqués de Valdecilla, Instituto de Formación e Investigación Marqués de Valdecilla, University of Cantabria, Santander, Spain. [241] Centre for Clinical Research in Neuropsychiatry, The University of Western Australia, Perth, Western Australia, Australia. [242] Western Australian Institute for Medical Research, The University of Western Australia, Perth, Western Australia, Australia. [243] Department of Psychiatry, Academic Medical Center, University of Amsterdam, Amsterdam, The Netherlands. [244] Department of Psychiatry, University Medical Center Groningen, University of Groningen, The Netherlands.